\documentclass[fleqn,usenatbib,useAMS]{mnras}

\usepackage{graphicx}	
\usepackage{amsmath}	
\usepackage{amssymb}	
\usepackage{multicol}        
\usepackage{bm}		
\usepackage{pdflscape}	
\usepackage{xcolor}

\usepackage[T1]{fontenc}
\usepackage{ae,aecompl}

\usepackage{newtxtext,newtxmath}

\title[The sub-galactic matter-power spectrum]{Probing sub-galactic mass structure with the power spectrum \\ of surface-brightness anomalies in high-resolution observations \\ of galaxy-galaxy strong gravitational lenses. II.~Observational constraints on the sub-galactic matter power spectrum}

\author[D. Bayer et al.]{D. Bayer,$^{1,}$$^{2,}$$^{3}$\thanks{Contact e-mail: \href{mailto:dbayer@swin.edu.au}{dbayer@swin.edu.au}}
S. Chatterjee,$^{3}$
L. V. E. Koopmans,$^{3}$
S. Vegetti,$^{4}$ 
J. P. McKean,$^{3,}$$^{5}$
\newauthor T. Treu,$^{6}$, C. D. Fassnacht$^{7}$ and  K. Glazebrook$^{1,}$$^{2}$   
\\
$^{1}$Centre for Astrophysics \& Supercomputing, Swinburne University of Technology, Hawthorn, VIC 3122, Australia\\
$^{2}$ARC Centre of Excellence for All Sky Astrophysics in 3 Dimensions (ASTRO 3D), Australia\\
$^{3}$Kapteyn Astronomical Institute, University of Groningen, PO Box 800, 9700 AV Groningen, the Netherlands\\
$^{4}$Max Planck Institute for Astrophysics, Karl-Schwarzschild-Strasse 1, D-85740 Garching, Germany\\
$^{5}$ASTRON, Netherlands Institute for Radio Astronomy, Postbus 2, NL-7990 AA, Dwingeloo, the Netherlands\\
$^{6}$Department of Physics and Astronomy, UCLA, 430 Portola Plaza, Los Angeles, CA 90095-1547, USA\\
$^{7}$Department of Physics and Astronomy, University of California, Davis, 1 Shields Ave. Davis, CA 95616, USA}

\date{Accepted XXX. Received YYY; in original form ZZZ}

\pubyear{2023}

\begin{document}
\label{firstpage}
\pagerange{\pageref{firstpage}--\pageref{lastpage}}
\maketitle

\begin{abstract}
Stringent observational constraints on the sub-galactic matter power spectrum would allow one to distinguish between the concordance $\Lambda$CDM and the various alternative dark-matter models that predict significantly different properties of mass structure in galactic haloes. Galaxy-galaxy strong gravitational lensing provides a unique opportunity to probe the sub-galactic mass structure in lens galaxies beyond the Local Group. Here, we demonstrate the first application of a novel methodology to observationally constrain the sub-galactic matter power spectrum in the inner regions of massive elliptical lens galaxies on 1-10 kpc scales from the power spectrum of surface-brightness anomalies in highly magnified galaxy-scale Einstein rings and gravitational arcs. The pilot application of our approach to Hubble Space Telescope (HST/WFC3/F390W) observations of the SLACS lens system SDSS~J0252+0039 allows us to place the following observational constraints (at the 99 per cent confidence level) on the dimensionless convergence power spectrum $\Delta^{2}_{\delta\kappa}$ and the standard deviation in the aperture mass $\sigma_{\rm AM}$: $\Delta^{2}_{\delta\kappa}<1$ ($\sigma_{\rm AM}< 0.8 \times 10^8 M_\odot$) on 0.5-kpc scale, $\Delta^{2}_{\delta\kappa}<0.1$ ($\sigma_{\rm AM}< 1 \times 10^8 M_\odot$) on 1-kpc scale and $\Delta^{2}_{\delta\kappa}<0.01$ ($\sigma_{\rm AM}< 3 \times 10^8 M_\odot$) on 3-kpc scale. These first upper-limit constraints still considerably exceed the estimated effect of CDM subhaloes. However, future analysis of a larger sample of galaxy-galaxy strong lens systems can substantially narrow down these limits and possibly rule out dark-matter models that predict a significantly higher level of density fluctuations on the critical sub-galactic scales.
\end{abstract}




\section{Introduction}
\label{Section:Introduction}

The dark-energy-plus-cold-dark-matter ($\Lambda$CDM) concordance cosmological model is known to successfully reproduce the observed large-scale (larger than $\sim1$  Mpc) distribution of matter in the Universe \citep[e.g.][]{Vogelsberger_Nature, EAGLE, Guo2016, Planck2018}. However, on smaller galactic and sub-galactic scales, theory and observations appear to diverge \citep[see][for a recent review on the small-scale challenges to the $\Lambda$CDM paradigm]{Bullock2017_Review}. One of the main discrepancies, known as the Missing Satellites Problem (MSP), lies in the fact that the number of dwarf satellite galaxies observed in the Local Group \citep[$\sim100$, e.g.][]{Local_Group_Satellites, DES_MWSatellites} is much lower than the numerous abundance of subhaloes populating galactic haloes in $\Lambda$CDM-based numerical simulations of the cosmological structure-formation process \citep[e.g.][]{Klypin,Moore,Diemand_Kuhlen_Madau_2007, MSP_3D, MSP_LMG, Bullock2017_Review, MSP_beyond_LG, 2019Fielder_MSP, 2019Navarro_Dark_Minihaloes}. 

The currently favoured interpretation of the MSP is that the missing subhaloes do exist but are extremely inefficient at forming stars due to a variety of baryonic processes, such as feedback from massive stars and active galactic nuclei, tidal stripping or photo-ionization squelching \citep[e.g.][]{Photoionization, photoionizingUV, Somerville, sawala2014, No_MSP, Despali_LOS}, and thus remain undetectable for conventional imaging surveys. Alternatively, the MSP might point towards alternative dark-matter models with higher thermal velocities in the early Universe, referred to as warm dark matter \citep[WDM,][]{WDMsim}, which predict a suppression of structure formation on the sub-galactic scales \citep[see e.g.][for some recent studies]{Menci2012, Nierenberg2013, Viel_PS_Lyman_alpha,Lovell2014, Vegetti2018, Zavala_Frenk, Hsueh2020WDM}. Lastly, the Local Group might just be a biased environment with less abundant substructure and, thus, not representative of the entire Universe \citep[e.g.][]{Plane_Satellite_Science}. Therefore, in order to fully clarify these ambiguities and test the aforementioned solutions to the small-scale tensions of the $\Lambda$CDM paradigm, it is crucial to constrain the properties of substructures in galaxies beyond the Local Universe on mass scales below that of luminous satellites.

The key techniques to search for the faint, or even truly dark, substructures in galaxies at cosmological distances are based on the study of their gravitational imprints on the lensed images of galaxy-scale strong gravitational lens systems. Substructures, if present in the foreground lens galaxy (or along its line-of-sight), introduce perturbations to the otherwise smooth lensing potential, thus leading to a deviation between the observed surface-brightness distribution of the lensed images and the prediction from the best-fitting smooth-lens model, e.g. in the form of flux-ratio anomalies in multiply imaged gravitationally lensed quasars \citep[e.g.][]{Mao, Metcalf_Madau, Dalal_Kochanek, Nierenberg2014_detection, Birrer_2keV, GilmanDM, Gilman_WDM_chills, Hsueh2020WDM} or surface-brightness anomalies in the lensed emission of an extended background galaxy \citep[e.g.][]{Blandford, Koopmans2005, VegettiKoopmans2009, Vegetti_2010_dwarf, Vegetti_DR_substructure, Rau2013, Vegetti2014, 2018Vegetti_Sterile_N, Ritondale}. Such anomalies can be measured, modelled and traced back to the underlying substructures in the lens galaxy. 

In particular, the \emph{gravitational-imaging technique} developed by \cite{Koopmans2005} and \citet{VegettiKoopmans2009} aims at the detection of individual massive dark-matter subhaloes in (massive elliptical) lens galaxies at cosmological distances, based on sophisticated modelling of surface-brightness anomalies measured in deep high-resolution imaging of galaxy-galaxy strong gravitational lens systems. The detection of two dark-matter subhaloes with masses $3.5 \times 10^9 M_\odot$ and $1.9 \times 10^8 M_\odot$ in lens galaxies at the redshift $z = 0.2$ and $z = 0.9$, respectively, reported in \cite{Vegetti_DR_substructure,Vegetti_Nature}, demonstrates that galactic subhaloes can be successfully identified as localised corrections to a smooth gravitational potential, with the detection threshold depending on the angular resolution of the available data. Recently, \cite{Despali_LOS} showed that this technique can also be used to detect individual line-of-sight haloes and \cite{Vegetti2018} derived the corresponding constraints on the properties of dark matter. However, despite these encouraging results, the current mass-detection threshold of the gravitational-imaging technique \citep[$\sim 10^8 M_\odot$ for HST-observations of SLACS lenses;][]{Vegetti2014, 2018Vegetti_Sterile_N} still lies above the mass regime in which the alternative dark-matter models could be easily distinguished. 

In an attempt to detect gravitational signatures of dark-matter subhaloes with masses below this detection threshold, an alternative statistical approach has emerged in the literature over the recent years \citep{Sander_thesis,Hezaveh_PS, Diaz_Rivero_2018, Saikat, ETHOS_Rivero, Beyond_subhaloes}. Within this framework, the entire population of low-mass subhaloes in the lens galaxy is modelled in terms of the sub-galactic matter power spectrum. This can potentially be constrained from the statistical analysis of the collectively-induced surface-brightness anomalies in the extended lensed images of the background source galaxy. Despite the considerable number of recent theoretical studies \citep{Sander_thesis,Hezaveh_PS, Diaz_Rivero_2018, Saikat, ETHOS_Rivero, Beyond_subhaloes}, there has not been a single attempt to apply this statistical approach to real observational data. This paper is part of a series aiming to close this gap. 

In the companion Paper I, we have introduced the methodology to measure the power spectrum of the hypothetical surface-brightness anomalies in high-resolution Hubble Space Telescope (HST) observations of galaxy-galaxy strong gravitational lens systems and applied it to a sub-sample of lens systems from the Sloan Lens ACS Survey \citep[SLACS,][]{Bolton2008}. In the present study, our aim is to extend this methodology and relate the power-spectrum measurement to the statistical properties of the underlying low-mass structures in the lens galaxy. The main challenge when applying this statistical approach to observed data, however, is that in reality the investigated surface-brightness anomalies may arise not only from the presence of the population of low-mass dark-matter subhaloes in the lens galaxy \citep[or field haloes along its line of sight, see e.g. ][]{Despali_LOS}, but also from a variety of different inhomogeneities in the total (dark and baryonic) projected mass density of the lens galaxy, such as e.g. small-scale density fluctuations in the distribution of dark matter, globular clusters, tidal streams, dynamical distortions in the baryonic mass distribution or edge-on discs \citep{Vegetti2014,Gilman,Hsueh2016, Hsueh_disk, Hsueh2017}, which are not explicitly included in the smooth-lens model. 

Thus, instead of considering the subhaloes only, we set out to constrain the statistical properties of the \emph{overall} density fluctuations in the total mass distribution of the lens galaxy (i.e. deviations from the best-fitting smooth-lens model) on the sub-galactic 1-10 kpc scales, modelled in terms of Gaussian random field (GRF) potential perturbations \citep{Sander_thesis, Hezaveh_PS, Saikat}. As a proof of concept, we apply the complete methodology to high-resolution HST/WFC3/F390W observations of the SLACS lens system SDSS~J0252+0039 \citep{UV_HST_proposal}. This analysis leads to the first-ever observational constraints on the power spectrum of GRF potential perturbations in a (massive elliptical) lens galaxy and the corresponding sub-galactic matter power spectrum. Future work will compare these constraints with predictions from hydrodynamical simulations, which incorporate both dark and baryonic matter \citep[such as Illustris or EAGLE,][]{Vogelsberger_Nature, EAGLE}.

The present paper is structured as follows. Section \ref{Section:Methodology} provides a concise description of the GRF-formalism applied to model the small-scale density fluctuations in the lens galaxy and presents our procedure adopted to uncover, quantify and interpret the resulting the surface-brightness anomalies in the lensed images. In Section \ref{Section:data_reduction_J0252}, we present the analysed HST observations of our pilot lens system SDSS J0252+0039. Sections~\ref{Section:GALFIT_J0252},~\ref{Section:Lens_Model_J0252} and~\ref{Section:HST_power_spectrum_J0252} describe the lens-galaxy subtraction, smooth lens modelling and the power-spectrum analysis of the residual surface-brightness fluctuations, respectively. In Section~\ref{Section:mock_perturbed_lenses_J0252}, we generate a catalogue of mock perturbed lensed images to be compared with the real observations. The inferred observational upper-limit constraints on the (projected) sub-galactic matter power spectrum are presented in Section \ref{Section:Results} and compared to predictions from the $\Lambda$CDM model in Section \ref{Section:Discussion}. The final Section \ref{Section:Conclusions} provides conclusions and implications of this work for further research.

For a consistent comparison of the lens models with earlier studies by \cite{Vegetti2014}, throughout this work we assume the following cosmology: $H_0 = 73 \ \mathrm{km \  s^{-1} Mpc^{-1}}$, $\Omega_M = 0.25$ and $\Omega_\Lambda = 0.75$. Given this cosmology, 1~arcsec corresponds to $4.11$ kpc at the redshift of the lens galaxy ($z_{L} = 0.280$) and $7.88$ kpc at the redshift of the source galaxy ($z_S = 0.982$).

\section{Methodology}
\label{Section:Methodology}

Low-mass structures in lens galaxies are commonly modelled in terms of corrections to the best-fitting smoothly-varying parametric model of the lensing potential $\psi_{0}( \textbf{\textit{x}})$ \citep[e.g.][]{Koopmans2005,VegettiKoopmans2009}. While previous studies have mainly focused on the detection of individual massive dark matter subhaloes \citep[e.g.][]{Vegetti_Nature, Vegetti2014} and line-of-sight haloes \citep{Despali_LOS}, here we investigate the overall departures from a smooth-lens model, i.e. small-scale density fluctuations in the total mass of the lens galaxy arising not only from subhaloes or haloes along the line-of-sight, but also from the complex (dark and baryonic) mass distribution of real lens galaxies in general. 

In the companion Paper I, we have introduced the methodology allowing us to measure the power spectrum of surface-brightness anomalies in high resolution HST-observations of galaxy-galaxy strong gravitational lens systems. In this paper, we extend this methodology and relate the measured power spectrum to the statistical properties of the underlying small-scale density fluctuations in the mass distribution of the lens galaxy. We model these as Gaussian random field (GRF) potential perturbations superposed on the best-fitting smoothly-varying lensing potential, as first suggested by \cite{Sander_thesis}, \cite{Hezaveh_PS} and \cite{Saikat}. 

In Section \ref{Section:GRF}, we summarise the concept of the GRF potential perturbations in the lens galaxy and the resulting surface-brightness anomalies in the lensed images. Section \ref{Section:Analysis overview} outlines the complete procedure to measure the power spectrum of the collectively-induced surface-brightness anomalies in the lensed images, relate it to the underlying potential perturbations in the lens galaxy and, finally, infer constraints on the sub-galactic matter power spectrum.

\subsection{GRF potential perturbations in the lens galaxy}
\label{Section:GRF}

We treat the small-scale density fluctuations in the lens galaxy as a statistical ensemble and model the associated potential corrections in terms of a homogeneous and isotropic Gaussian potential perturbation field $\delta\psi_{\rm{GRF}}( \textbf{\textit{x}})$ (with $\langle \delta\psi_{\rm{GRF}}(\textbf{\textit{x}})\rangle=0$) superposed on a smoothly-varying lensing potential $\psi_{0}( \textbf{\textit{x}})$ \citep{Sander_thesis, Hezaveh_PS, Saikat, Knotty_Lenser_Giorgos}. More specifically, we assume that the lensing potential $\psi( \textbf{\textit{x}})$ of a lens galaxy can to the first order be approximated by the sum of a smoothly-varying parametric component $\psi_{0}(\textbf{\textit{x}})$ and a Gaussian potential perturbation field $\delta\psi_{\rm{GRF}}(\textbf{\textit{x}})$: 
\begin{equation}
\psi( \textbf{\textit{x}})\approx \psi_{0}( \textbf{\textit{x}})+\delta \psi_{\rm{GRF}}( \textbf{\textit{x}}),
\label{psi_approx_J0252}
\end{equation}
with no covariance between $\delta\psi_{\rm{GRF}}( \textbf{\textit{x}})$ and $\psi_{0}( \textbf{\textit{x}})$.

In this approximation, the deflection angle \textbf{$\alpha(\textbf{\textit{x}})$} can be separated into the deflection caused by the smooth component of the lensing potential \textbf{$\alpha_{0}(\textbf{\textit{x}})$} and a perturbative deflection-angle field $\bm{\delta \alpha}_{\rm{GRF}}(\textbf{\textit{x}})$ caused by the differential lensing effect of the potential perturbations $\delta\psi_{\rm{GRF}}(\textbf{\textit{x}})$:
\begin{equation}
\bm{\alpha} (\textbf{\textit{x}})= \nabla \psi(\textbf{\textit{x}}) \approx \nabla \psi_{0}(\textbf{\textit{x}}) +  \nabla \delta\psi_{\rm{GRF}}(\textbf{\textit{x}}) = \bm{\alpha_{0}}(\textbf{\textit{x}}) + \bm{\delta \alpha}_{\rm{GRF}} (\textbf{\textit{x}}). 
\label{deflection_angle_total_J0252}
\end{equation}
The resulting differential deflection-angle field $\bm{\delta \alpha}_{\rm{GRF}}(\textbf{\textit{x}})$ can be in this case directly linked to the underlying potential perturbations $\delta\psi_{\rm{GRF}}(\textbf{\textit{x}})$ via:
\begin{equation}
\bm{\delta \alpha}_{\rm{GRF}}(\textbf{\textit{x}})= \nabla \delta\psi_{\rm{GRF}}(\textbf{\textit{x}}),
\label{deflection_angle_perturbed_J0252}
\end{equation}
independently of the smooth lensing component $\psi_{0}(\textbf{\textit{x}})$. Similarly, the associated convergence-perturbation field $\delta \kappa_{\rm{GRF}}( \textbf{\textit{x}})$ (i.e. surface-mass density perturbations in units of the critical surface-mass density for lensing) is related to the potential-perturbation field $\delta\psi_{\rm{GRF}}(\textbf{\textit{x}})$ via the Poisson equation:
\begin{equation}
\nabla^2 \delta \psi_{\rm{GRF}}(\textbf{\textit{x}}) = 2\times  \delta \kappa_{\rm{GRF}}( \textbf{\textit{x}}).
\label{Poisson_delta_J0252}
\end{equation}

A crucial feature of a Gaussian random field is that its properties are entirely characterised by the second-order statistics. Hence, the statistical behaviour of the Gaussian potential perturbations $\delta\psi(\textbf{\textit{x}})$ is fully described by the 2-point correlation function or, alternatively, its Fourier transform -- the power spectrum. Following \cite{Saikat}, we assume the power spectrum of $\delta\psi(\textbf{\textit{x}})$ to be isotropic and to follow a power law: 
\begin{equation}
P_{\delta\psi}(k) = A \times k^{-\beta}.
\label{eq:power_law_PS_J0252}
\end{equation} 
The amplitude $A$ is related to the total variance of the potential perturbations $\sigma^2_{\delta\psi}$ (integrated between the spatial scales equal to the inverse of the image length and the inverse of the pixel scale, respectively), while the power-law slope $\beta$ describes the distribution of this variance over the different length scales (i.e. $k$-modes) and, thus, determines the scale dependence of the hypothetical small-scale structure in the lens galaxy. In what follows, we refer to a specific combination of $\sigma^2_{\delta\psi}$ and $\beta$ as a \emph{matter-power-spectrum model}. We choose the convention in which the wavenumber $k$, measured in $\mathrm{arcsec^{-1}}$, corresponds to the reciprocal wavelength $\lambda$: 
\begin{equation} k\equiv\lambda^{-1} \label{eq:k_definition_J0252} \end{equation}
of the associated harmonic wave $e^{-2 \pi i \textit{k} \cdot \textit{x}}$ in the Fourier representation of the GRF field.
 
For a science image with a side length $L$ (measured in arcsec), we determine $A$ by specifying the overall variance of the potential perturbations $\sigma^2_{\delta\psi}\equiv\langle (\delta \psi - \langle \delta \psi \rangle )^2 \rangle = \langle \delta \psi^2 \rangle$ in the considered field-of-view: 
\begin{equation}
\int_{k_{x}}\int_{k_{y}} P_{\delta\psi} \Big(A,\beta,\sqrt{k_{x}^2+k_{y}^2} \Big) \ dk_{x} \ dk_{y}= \sigma^2_{\delta\psi},
\label{eq:normalisation_J0252}
\end{equation}
where the integration is performed over the corresponding Fourier grid and the wavenumbers $k_{x}$ and $k_{y}$ are calculated according to equation (\ref{eq:k_definition_J0252}). Substituting $P_{\delta\psi}(k)$ in equation (\ref{eq:normalisation_J0252}) with equation (\ref{eq:power_law_PS_J0252}) and replacing the integrals by a summation over discrete pixels with the size $dk_{x}=dk_{y} = L^{-1}$ in $k$-space finally leads to the following normalisation condition: 
\begin{equation}
A \Big( \sigma^2_{\delta\psi}, \beta, L \Big) = \frac{L^2 \sigma^2_{\delta\psi}}{\sum_{k_{x}}\sum_{k_{y}} \Big( \sqrt{k_{x}^2+k_{y}^2} \Big) ^{-\beta}}.
\label{eq:A_J0252}
\end{equation}
As can be seen from equation (\ref{eq:A_J0252}), this normalisation depends not only on the specific combination of $\beta$ and $\sigma^2_{\delta\psi}$, but also on the field-of-view and the pixel scale of the analysed image. 

The perturbative effect of such Gaussian potential fluctuations $\delta\psi_{\rm{GRF}}(\textbf{\textit{x}})$ in the lens galaxy leads to observable surface-brightness anomalies in the lensed images $\delta I_{\rm{GRF}}(\textbf{\textit{x}})$ of the background source galaxy, measured with respect to the smooth-lens model $I_0(\textbf{\textit{x}})$:
\begin{equation}
\begin{split}
& \delta I_{\rm{GRF}}(\textbf{\textit{x}}) = I_{\rm{GRF}}(\textbf{\textit{x}}) - I_0(\textbf{\textit{x}}) = \\
& = S\big( \textbf{\textit{x}} - \nabla \psi_0(\textbf{\textit{x}})-\nabla \delta\psi_{\rm{GRF}}(\textbf{\textit{x}})\big) - S\big(\textbf{\textit{x}} - \nabla \psi_0(\textbf{\textit{x}})\big)=\\
& =S\big( \textbf{\textit{x}} - \bm{\alpha_{0}}(\textbf{\textit{x}}) - \bm{\delta \alpha}_{\rm{GRF}} (\textbf{\textit{x}})\big) - S\big(\textbf{\textit{x}} -\bm{\alpha_{0}}(\textbf{\textit{x}}) \big).
\end{split}
\end{equation}
Using the methodology developed in this work, we seek to derive observational constraints on the variance $\sigma^2_{\delta\psi}$ and the slope $\beta$ of the power-law power spectrum $P_{\delta\psi}(k)$, as defined in equations (\ref{eq:power_law_PS_J0252}--\ref{eq:A_J0252}), from the power spectrum of surface-brightness anomalies $\delta I_{\rm{GRF}}(\textbf{\textit{x}})$ measured in high-resolution HST-observations of galaxy-galaxy strong gravitational lens systems. 

\subsection{Analysis overview}
\label{Section:Analysis overview}

Our approach to constrain the level of such small-scale density fluctuations in massive elliptical lens galaxies consists of three main components:

\begin{itemize}
\item First, we extract the surface-brightness anomalies from the lensed images of an observed lens system and quantify their statistical properties in terms of the power spectrum $P_{\delta I}(k)$, see Paper I for a thorough discussion of the methodology and the modelling degeneracies. To this end, we perform the following steps:
\begin{enumerate}

\item HST-observations and data reduction by means of the \textsc{drizzlepac} package \citep{Gonzaga2012}, see Section \ref{Section:data_reduction_J0252}; \vspace{0.1cm}

\item modelling and subtraction of the surface-brightness contribution from the foreground lens galaxy using \textsc{galfit} \citep{GALFIT_paper}, see Section \ref{Section:GALFIT_J0252}; \vspace{0.1cm}

\item simultaneous reconstruction of the smooth lensing potential $\psi_{0}( \textbf{\textit{x}})$ and the intrinsic surface-brightness distribution of the background source galaxy $S(\textbf{\textit{y}})$, using the adaptive and grid-based Bayesian lens-modelling technique by \cite{VegettiKoopmans2009}, see Section \ref{Section:Lens_Model_J0252};\vspace{0.1cm}

\item statistical quantification of the residual surface-brightness fluctuations in the lensed images in terms of the azimuthally-averaged power spectrum $P_{\delta I}(k)$, see Section \ref{Section:HST_power_spectrum_residuals}; \vspace{0.1cm}

\item estimation and correction for the noise power spectrum, including both the sky background and the flux-dependent photon shot noise, to obtain the power spectrum of surface-brightness anomalies in the observed lensed images; see Section \ref{Section:Noise power spectrum}.\vspace{0.1cm}

\end{enumerate}

\vspace{0.1cm}

To be conservative in coping with the degeneracies discussed in Paper I, we treat any residual surface-brightness fluctuations remaining in the lensed images after the lens-galaxy subtraction, the smooth-lens modelling and the noise correction as an upper limit to the effect caused by the density fluctuations in the lens galaxy.

\item Second, we generate a catalogue of mock lensed images perturbed by GRF potential fluctuations $\delta\psi_{GRF}(\textbf{\textit{x}})$ with known statistical properties (i.e. GRF power spectrum $P_{\delta\psi}(k)$ with the total variance $\sigma^2_{\delta\psi}$ and the power-law slope $\beta$, as defined in equations \ref{eq:power_law_PS_J0252}-\ref{eq:A_J0252}) for a systematic comparison with the observed data. Subsequently, we subtract the smooth lens model obtained for the data from the mocks and quantify the resulting residual images in terms of the power spectrum; see Section \ref{Section:mock_perturbed_lenses_J0252}.

\item Third, we compare the obtained power spectra of surface-brightness anomalies in the mock perturbed lensed images to the upper limit inferred from the observational data and estimate the exclusion probability for each considered combination of $\sigma^2_{\delta\psi}$ and $\beta$, given the observational measurement. Based on these results, we additionally derive upper-limit constraints on the power spectrum of  perturbations in the deflection angle $P_{\delta\alpha}(k)$ and the convergence (i.e. projected mass density) $P_{\delta\kappa}(k)$; see Section \ref{Section:Results}. 
\end{itemize}

We emphasize that we intend to apply our method solely to lens systems for which the effect of single dominant subhaloes and satellite galaxies has already been modelled in previous studies. Moreover, in the future, we plan to additionally use hydrodynamical simulations to thoroughly test the validity of the Gaussianity assumption for the hypothetical departures of the total mass distribution in massive elliptical (lens) galaxies from a smooth model.

\section{Observations and data reduction} 
\label{Section:data_reduction_J0252}

The mass of the smallest structures that can be recovered in a galaxy-scale halo using the method of gravitational imaging \citep{Koopmans2005,VegettiKoopmans2009} is limited by the sensitivity of the observational setup to tiny surface-brightness perturbations in the lensed images. As argued by \cite{Blandford}, \cite{Koopmans2005} and explicitly demonstrated by \cite{Rau2013}, the surface-brightness anomalies induced in the lensed images by a given density fluctuation in the lens galaxy are enhanced when the lensed background source is either compact or highly structured. Taking into account that blue star-forming regions are considerably more structured than the redder old stellar populations, the choice of a lens system with a star-forming source galaxy combined with the selection of an ultra-violet observational filter might substantially improve the overall sensitivity of our approach. 

Following this idea, we apply our methodology to HST/WFC3/F390W-imaging data (with the central wavelength of 390 nm, referred to as \textit{U-band} in the remainder of the paper) of the galaxy-galaxy strong gravitational lens system SDSS J0252+0039 from the SLACS Survey. This lens system consists of a massive elliptical lens galaxy at redshift $z_{\rm l} = 0.280$ and a blue star-forming source galaxy at redshift ${z_{\rm s}} = 0.982$ \citep{Bolton2008}.
The choice of SDSS J0252+0039 is motivated by the large mass of its lens galaxy (total mass of $10^{11.25} M_\odot$ inside the Einstein radius), the high surface-brightness gradient of the star-forming background galaxy and the relatively simple lens geometry which reduces the uncertainty in the lens modelling \citep{Auger2009}.

As one of the SLACS gravitational lens systems, SDSS J0252+0039 has already been modelled, based on HST-observations at near-infrared (in F160W: Program 11202, PI Koopmans) and optical wavelengths (in F814W: Program 10866, PI Bolton and in F606W: Program 11202, PI Koopmans), see \cite{Bolton2008} and \cite{Auger2009, Auger2010}. In particular, \cite{Vegetti2014} applied the technique of gravitational imaging \citep{Koopmans2005, VegettiKoopmans2009} to the HST/ACS data in the V- and I-bands to search for signatures of individual massive sub-haloes in the lens galaxy. No gravitational signatures of localised substructure have been identified above the mass-detection threshold of $\sim 10^8 M_\odot$. Here we take advantage of the enhanced source gradient in the U-band observations and search for the collectively-induced gravitational imprints of much smaller density fluctuations.

\begin{figure}
\includegraphics[width=\columnwidth]{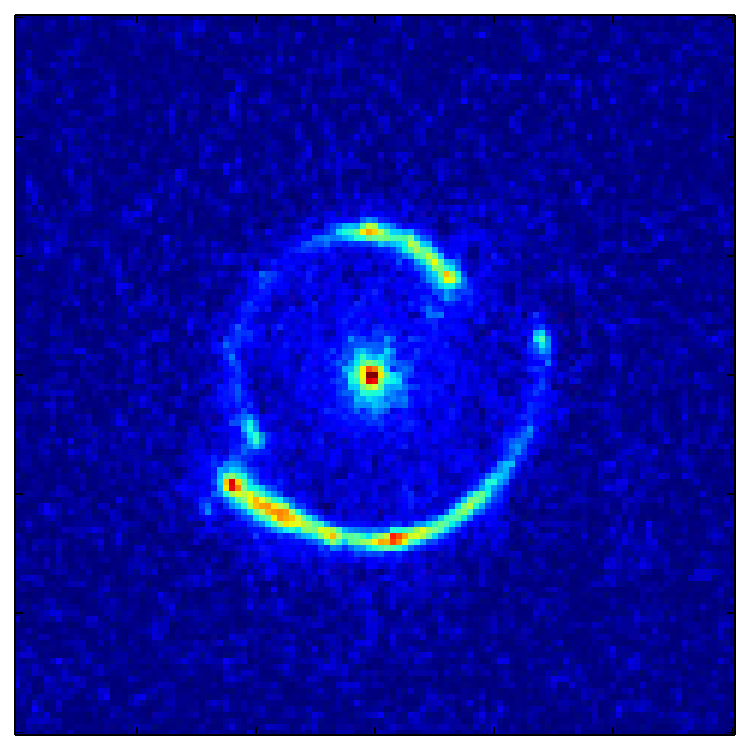}
\caption{HST/WFC3/F390W-imaging of the strong gravitational lens system SDSS J0252+0039: the drizzled science image with the side length of 4.48 arcsec.}
\label{fig:HST_imaging_J0252}
\end{figure}

The HST/WFC3/F390W-observations of SDSS J0252+0039 were carried out on August 25, 2013 \citep[Program 12898, PI Koopmans,][]{UV_HST_proposal}. We retrieve the acquired eight dithered flat-field calibrated images from the MAST archive\footnote{http://archive.stsci.edu/hst/search.php} and apply the drizzle method \citep{Gonzaga2012} to combine them into the final science image. For this, we use the \textsc{astrodrizzle} task from the \textsc{drizzlepac} package in its default configuration, with the original HST/WFC3 rotation, the output-pixel size of 0.0396 arcsec and the drop size equal to the original pixel size (\textit{final pixfrac} = 1). Our analysis is based on an image cutout of 121 by 121 pixels (corresponding to an area of 4.48 arcsec on a side) centered on the brightest pixel of the lens galaxy, which we present in Fig.~\ref{fig:HST_imaging_J0252}.

A known effect of the drizzling procedure are the correlations between adjacent pixels in the output science image \citep[see] []{Casertano}. These correlations are additionally enhanced by the effect of the charge-transfer inefficiency (CTI) in the HST/WFC3/UVIS-CCDs \citep[see e.g.][]{Baggett_CTE}. To investigate and possibly alleviate these instrumental effects, we repeat the drizzling procedure with different settings (i.e. different values of the \textit{final pixfrac} parameter and the output pixel scale) and compare the results to those obtained for the default configuration. Moreover, we perform a comparison between the power spectra obtained by drizzling either the original or the CTI-corrected flat-field calibrated exposures for a sample of blank-sky cutouts (located in vicinity to the lens system). Based on this analysis, discussed in detail in Paper I, we decide to proceed by drizzling the original dithered exposures in the default configuration of the \textsc{astrodrizzle} task.

To account for further observational effects present in the science image, we obtain the point-spread function (PSF) of the HST/WFC3/UVIS optics using the PSF-modelling software \textsc{tinytim}\footnote{http://www.stsci.edu/hst/observatory/focus/TinyTim} \citep{TinyTim_Paper}, with the approximation of the G8V spectral type for the lens galaxy \citep[based on its known magnitudes in the V-, I- and H-bands from][]{Auger2009}. Even though the TinyTim-PSF might not be a perfect representation of the real telescope optics, any minor deviations from the true PSF and the assumed spectral-energy distribution of the lens system affect the measured power spectra of the residual surface-brightness fluctuations only on scales below the full width at half maximum ($\mathrm{FWHM}$) of the PSF, corresponding to wavenumbers $k \gtrapprox \mathrm{FWHM}^{-1}={(0.07 \ \mathrm{arcsec})}^{-1}\approx 14 \ \mathrm{arcsec^{-1}}$, which is beyond the regime considered in this work (see Section \ref{Section:Noise power spectrum}).

\begin{figure}
\includegraphics[scale=1.93]{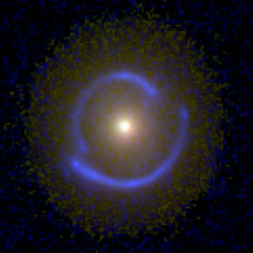}
 \caption{Colour-composite image of the strong gravitational lens system SDSS J0252+0039 combining the new HST/WFC3/F390W-photometry (\textit{blue}) with the already existing HST-observations in the visual (F814W; \textit{green}) and infrared (F160W; \textit{red}) bands, obtained using the \textsc{stiff} software.}
 \label{fig:HST_combined}
\end{figure}

Finally, for visual purposes and in order to assess the possible presence of dust in the studied lens system, we use the \textsc{stiff}\footnote{https://www.astromatic.net/software/stiff} package to create a colour-composite image of SDSS J0252+0039. The resulting Fig.~\ref{fig:HST_combined} shows a smooth early-type lens galaxy and uniformly blue lensed images of the background galaxy. A visual inspection of this image does not identify any indications of dust extinction either in the lens or the source galaxy. This conclusion is confirmed by a more quantitative dust analysis, discussed in Appendix \ref{Appendix:Dust}.

\section{Lens-galaxy subtraction} 
\label{Section:GALFIT_J0252}

The Einstein radius of SDSS J0252+0039 ($\sim$ 1~arcsec) is comparable to the effective radius of its massive elliptical lens galaxy. Consequently, the lensed images overlap with the inner region of the lens galaxy, as can be seen in Fig.~\ref{fig:HST_imaging_J0252}. To correct for the light contribution from the lens galaxy, we model its surface-brightness distribution with a combination of several S\'ersic components \citep{Sersic1963} -- a parametric class of smooth light profiles widely used to model early-type galaxies.

\begin{figure*}
\centering
\includegraphics{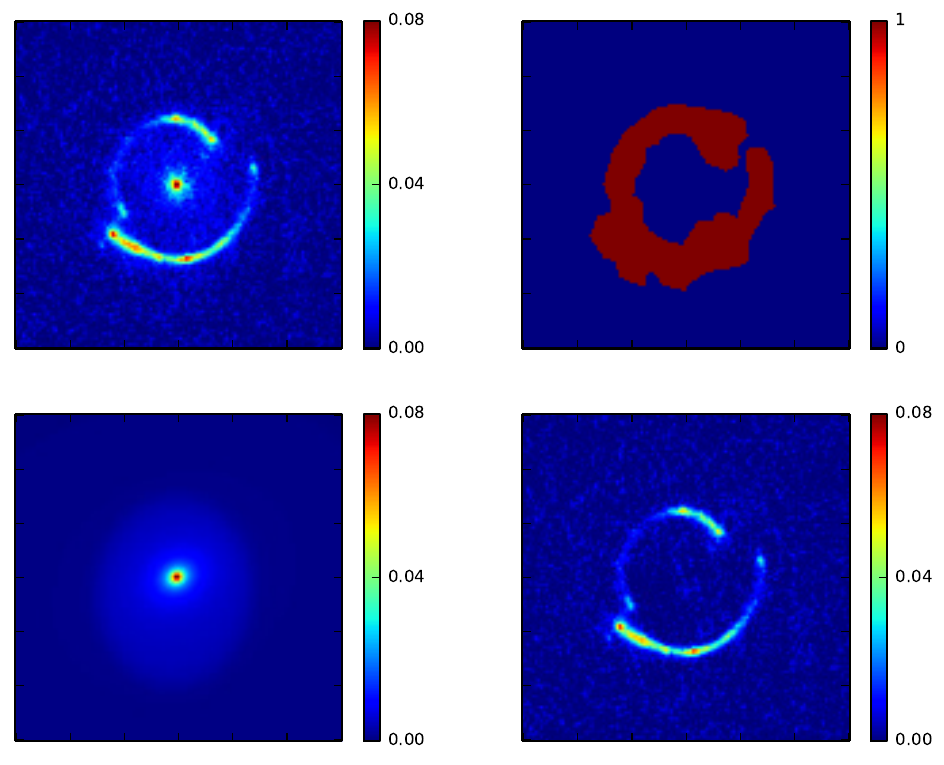}
\caption{Lens-galaxy subtraction for SDSS J0252+0039 using \textsc{galfit}. \textit{Top row}: the drizzled science image in F390W \textit{(left panel)} and the applied mask, covering the lensed source emission \textit{(right panel)}. \textit{Bottom row}: the best-fitting \textsc{galfit} model of the surface-brightness distribution in the lens galaxy \textit{(left panel)} and the lens-galaxy-subtracted residual image \textit{(right panel)}.}
\label{fig:Galfit_J0252}
\end{figure*}

The applied lens-galaxy-subtraction procedure is illustrated in Fig.~\ref{fig:Galfit_J0252}. We perform the modelling by means of the two-dimensional fitting algorithm \textsc{galfit} \citep{GALFIT_paper} after masking out all pixels covering the lensed images, as shown in the top row. The best-fitting \textsc{galfit} model, shown in the bottom left panel of Fig~\ref{fig:Galfit_J0252}, consists of two S\'ersic components with S\'ersic indices $n = 4.44$ and $n = 0.09$ and effective radii 1.2 and 0.96 arcsec (corresponding to $\sim4.8$ and $\sim3.8$ kpc at the redshift of the lens), respectively. The small value of the second S\'ersic index suggests that the light distribution of the lens galaxy can be well described by the standard de Vaucouleurs profile (i.e. S\'ersic profile with $n=4$) and a diffuse stellar halo. As an alternative, we additionally carry out the lens-galaxy subtraction using the \textsc{b-spline} algorithm \citep{Bolton2006}, but choose to continue our analysis based on the \textsc{galfit} model due to the slightly higher Bayesian evidence of the resulting best-fitting smooth-lens model, see Section \ref{Section:Lens_Model_J0252}. The final lens-galaxy-subtracted image is presented in the bottom right panel of Fig~\ref{fig:Galfit_J0252}. This residual image constitutes our best estimate of the lensed emission from the background galaxy only and is as such used for the smooth-lens modelling. 

\section{Smooth lens modelling} 
\label{Section:Lens_Model_J0252}

In search of possible surface-brightness anomalies in the lensed images, we first model the studied lens system assuming that the global mass distribution in the lens galaxy is smooth and to first order well described by the power-law elliptical mass-distribution model \citep[PEMD,][]{Barkana1998} in an external shear field. This smooth lens modelling is performed by means of the adaptive and grid-based Bayesian lens-modelling technique of \cite{VegettiKoopmans2009} that allows us to find the best-fitting parameter values of the PEMD model for the lens galaxy and, simultaneously, reconstruct the unlensed surface-brightness distribution of the background galaxy on an adaptive grid in the source plane. We follow the parametrization used by \cite{VegettiKoopmans2009} and model the projected mass density of the lens galaxy in terms of the convergence $\kappa$:
\begin{equation}
\kappa(x,y) = \frac{b \ (2-\frac{\gamma}{2}) \ q^{\gamma-3/2}}{2(x^{2}q^{2}+y^{2})^{(\gamma-1)/2}}.
\label{eq:smooth_lens_convergence}
\end{equation} 
This is a function of the position in the lens plane $(x,y)$ and has the following parameters -- the lens strength $b$, the position angle of the major axis $\theta$ (with respect to the original telescope rotation), the (minor to major) axis ratio $q$, the (three-dimensional) mass-density slope $\gamma$ (with $\gamma=2$ for the isothermal case), the centre coordinates of the mass distribution in the lens plane $x_{0}$ and $y_{0}$, the external shear strength $\Gamma$ and its position angle $\Gamma_{\theta}$. This parametrization tries to reduce the degeneracy between the mass enclosed by the lensed images, the axis ratio of the lens galaxy and its mass-density slope.

Several options for the inversion of the lensing operation are available to choose from when using the smooth-lens-modelling code by \cite{VegettiKoopmans2009}. To prevent overfitting, the reconstruction of the background galaxy can be regularized by applying an adaptive or non-adaptive, variance, gradient or curvature source-plane regularization. Furthermore, one can choose the source-grid resolution, which is characterised by the fraction and spacing of pixels that are cast back from the image plane to the source plane when generating a grid for the source reconstruction. More specifically, the source-grid resolution is controlled by the parameter $n$ that sets the linear size of a square in the image plane out of which only one pixel is cast back to the source plane. For example, if $n = 3$ only one pixel out of each contiguous $3\times3$-pixel area is used to create the Delauney-tesselation grid in the source plane. We note that, even if $n>1$, \emph{all} pixels are still used to calculate the Bayesian evidence and compare the model to the data. 

As pointed out by \cite{Suyu2006_pixelized_inversion} and \cite{Vegetti2014}, the optimal choice for the regularization and the source-grid resolution depends crucially on the smoothness of the modelled lensed images and may be different for each specific lens system. Thus, the common practice is to perform the smooth lens modelling for different combinations of the above options, and find the optimal settings based on the highest Bayesian evidence. Our tests in this respect show that the chosen source-grid properties hardly affect the best-fitting parameter values of the lensing potential. We find, however, that they have a significant effect on the reconstructed unlensed surface-brightness distribution of the background source galaxy and, consequently, on the level of the residual surface-brightness fluctuations remaining in the lensed images after the best-fitting smooth-lens model has been subtracted.

\begin{figure*}
 \includegraphics[scale=0.68]{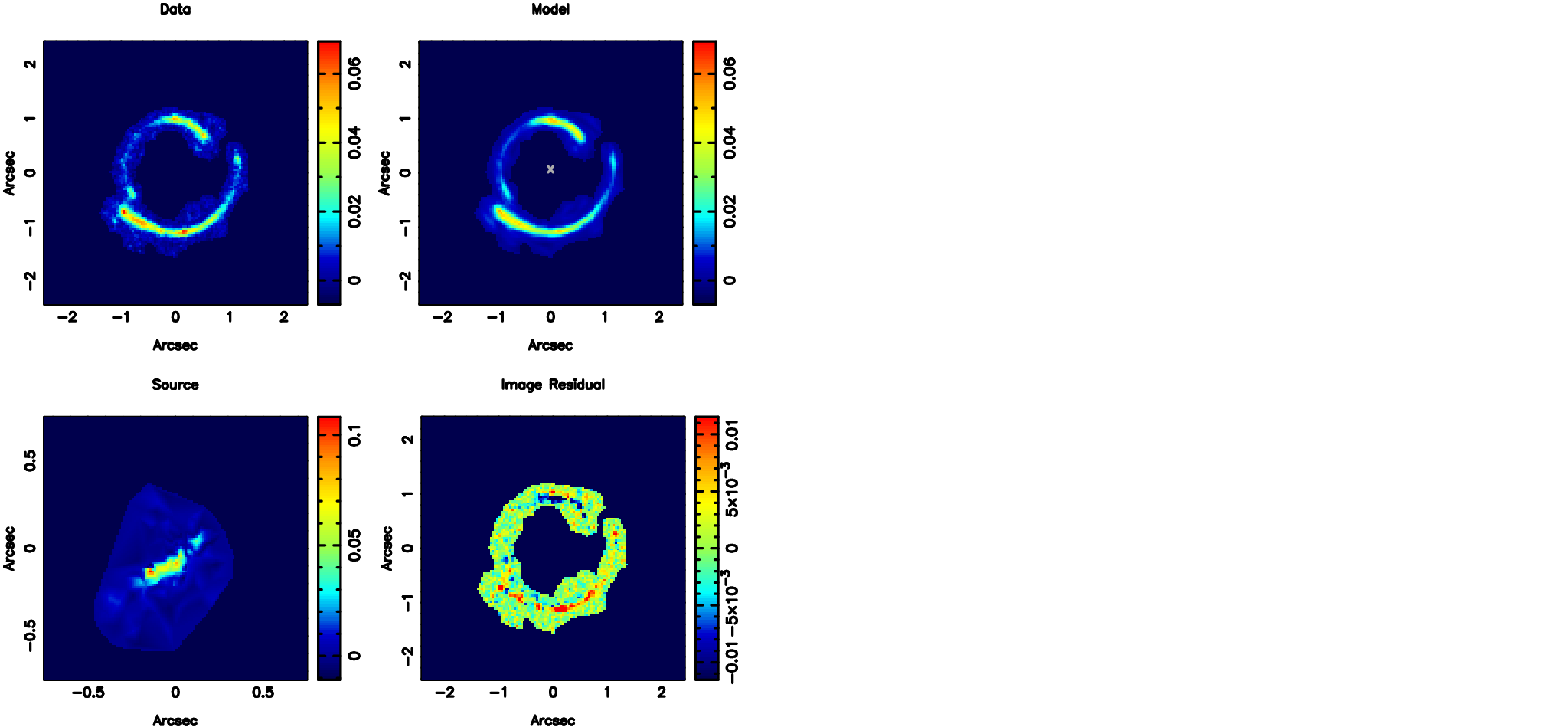}
 \centering
 \caption{Smooth lens modelling of SDSS J0252+0039 in the U-band (HST/WFC3/F390W) by means of the adaptive and grid-based Bayesian lens-modelling technique of \citet{VegettiKoopmans2009}, performed with the source-grid resolution $n=3$, i.e. casting back only one pixel out of each contiguous $3\times3$-pixel area from the image plane to the source plane. \textit{Top row}: the lens-galaxy-subtracted image overlaid with a mask, used as input for the smooth lens modelling \textit{(left panel)} and the best-fitting smooth-lens model of the lensed images \textit{(right panel)}. \textit{Bottom row}: the reconstructed intrinsic surface-brightness distribution of the background galaxy \textit{(left panel)} and the residual image showing the remaining surface-brightness fluctuations in the lensed images, possibly caused by small-scale mass structure in the lens galaxy \textit{(right panel)}.}
 \label{fig:smooth_lens_n3}
\end{figure*}

We achieve the highest value of the marginalized Bayesian evidence when the smooth lens modelling is performed by casting back each pixel from the lens plane to the source plane (referred to as $n = 1$) and using the adaptive gradient source-grid regularization, see Fig.~\ref{fig:smooth_lens_n3}. The best-fitting parameter values of the lensing potential are presented in Table~\ref{tab:smooth_lens_parameters_J0252}. These are in a good agreement with the parameter values inferred by \cite{Vegetti2014} in an earlier analysis of the HST/ACS/F814W-data (I-band), except for the apparent discrepancy in the position angle of the major axis $\theta$. This discrepancy can be explained by the rotational invariance owing to the nearly spherical symmetry of the modelled lens galaxy (axis ratio $q$ very close to 1) and the negligible external shear $\Gamma$, hardly altering the lensing potential. All in all, the best-fitting parameter values inferred in both bands indicate that SDSS J0252+0039 has a nearly spherical isothermal mass-density profile ($\gamma \approx 2$) with the Einstein radius of $\Theta_{E} \approx b \approx$ 1 arcsec. 

However, despite a remarkably good agreement between the model and data, the source reconstruction obtained using $n = 1$ turns out to be under-regularised, i.e. most of the surface-brightness fluctuations in the lensed images, and partially even the observational noise, are "absorbed" in the source structure. The power spectrum of the residual surface-brightness fluctuations (Section~\ref{Section:HST_power_spectrum_J0252}) remaining in the lensed images after the subtraction of the best-fitting smooth-lens model lies below the noise power spectrum for both adaptive and non-adaptive source-grid regularisation (see Paper I). While a thorough analysis of the degeneracy between the fluctuations in the lensing potential and the intrinsic source structure is the subject of a future work \citep[see also][for more details]{Saikat_thesis}, we continue our study based on the (more conservative) source reconstruction obtained using a lower grid resolution ($n = 3$), while keeping fixed the parameter values inferred from the highest-resolution model ($n = 1$), see Fig.~\ref{fig:smooth_lens_n3} and Table~\ref{tab:smooth_lens_parameters_J0252}. The particular choice of $n = 3$ is motivated by the results of our tests carried out for a mock lens system mimicking SDSS J0252+0039 (see Paper I). These tests suggest that the degeneracy between the density fluctuations in the lens galaxy and the intrinsic source structure is significantly reduced when the modelling is performed with $n = 3$ (or higher).

\begin{figure}
\centering
 \includegraphics[width=\columnwidth]{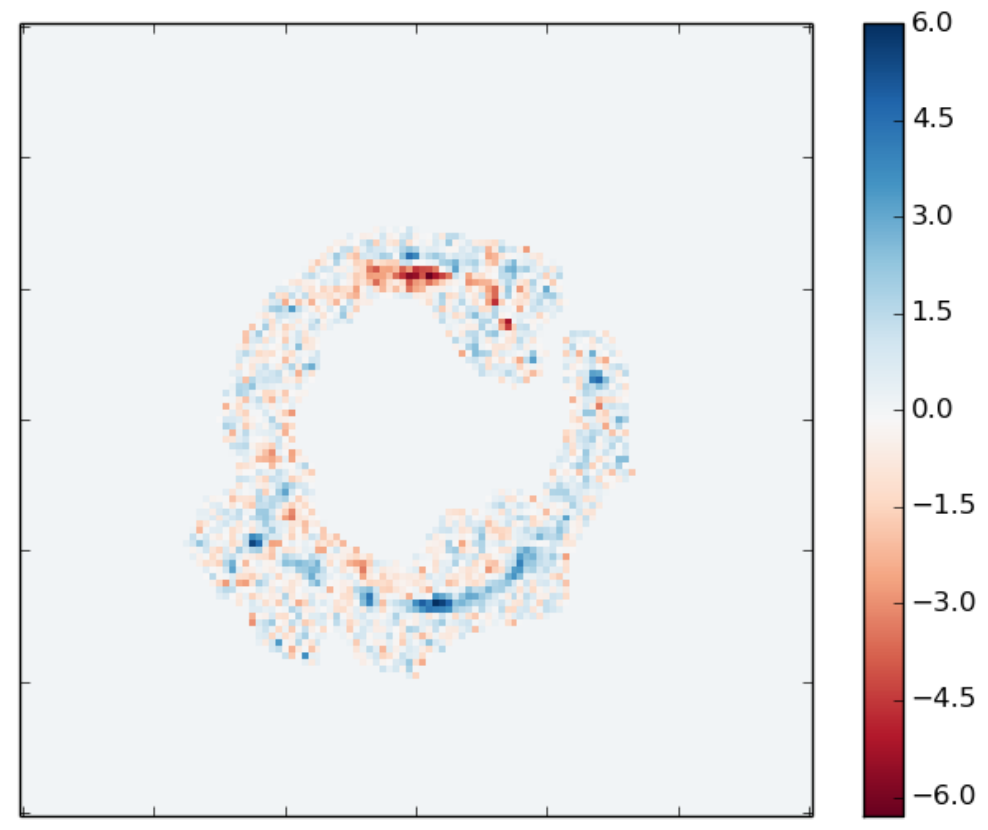}
 \caption{The signal-to-noise ratio of the residual surface-brightness fluctuations remaining in the lensed images of SDSS J0252+0039 after subtraction of the smooth-lens model obtained with the source-grid resolution of $n = 3$.}
 \label{fig:SN_residual_n3}
\end{figure}

The obtained model of the lensed images (top right panel of Fig.~\ref{fig:smooth_lens_n3}), which would be observed if the lensing potential was indeed smooth, is subsequently subtracted from the observed data to uncover residual surface-brightness fluctuations. These indicate a possible deviation of the true lensing potential from the assumed smooth (PEMD) model. As can be seen from the residual image in the bottom right panel of Fig.~\ref{fig:smooth_lens_n3} and the corresponding signal-to-noise ratio image presented in Fig.~\ref{fig:SN_residual_n3}, the revealed residual surface-brightness fluctuations significantly exceed the noise level.

\begin{table}
	\centering
	\caption{Parameter values of the best-fitting PEMD model for the lens galaxy SDSS J0252+0039 inferred based on the U-band (F390W) imaging, in comparison to the I-band (F814W) reconstruction by \citet{Vegetti2014}. The optimized parameters are the lens strength $b$, the position angle $\theta$ (with respect to the original telescope rotation), the axis ratio $q$, the (three-dimensional) mass-density slope $\gamma$, the external shear strength $\Gamma$ and its position angle $\Gamma_{\theta}$. In both bands, the reconstruction was carried out using the adaptive gradient source regularization and casting back each pixel ($n = 1$). The typical statistical errors of our U-band reconstruction are of the order  $10^{-1}$ for the angles and $10^{-3}$ for the remaining parameters.}
	\label{tab:smooth_lens_parameters_J0252}
	\begin{tabular}{ccccccc} 
		\hline
		Filter & $b$ [\arcsec] & $\theta$ [deg.] & $q$ & $\gamma$ & $\Gamma$ & $\Gamma_{\theta}$ [deg.]\\
		\hline
		F390W & 0.996 & 150.1 & 0.978 & 2.066 & -0.015 & 81.4\\
		F814W & 1.022 & 26.2 & 0.943 & 2.047 & 0.009 & 101.8\\
		\hline
	\end{tabular}
\end{table}

\section{Power-spectrum analysis of the surface-brightness anomalies in the lensed images}
\label{Section:HST_power_spectrum_J0252}

In this section, we estimate the power spectrum of surface-brightness anomalies in SDSS J0252+0039, defined as residual surface-brightness fluctuations in the lensed images caused by the presence of density fluctuations in the lens galaxy (i.e. deviations of the real mass distribution from the best-fitting PEMD model). To start with, in Section \ref{Section:HST_power_spectrum_residuals}, we take into account all the residual surface-brightness fluctuations revealed in the lensed images after the subtraction of the lens-galaxy light and the best-fitting smooth-lens model. We note, however, that such residual fluctuations might originate not only from the possible mass structure in the lens galaxy, but also from other phenomena, for example the observational noise, uncertainties in the PSF-model or intrinsic structure in the lensed source that has not been recovered in the lens-modelling procedure due to the choice of a relatively low source-grid resolution ($n = 3$). Hence, in order to constrain the surface-brightness anomalies resulting solely from the density fluctuations in the mass distribution of the lens galaxy, it is crucial to quantify the other effects, in particular the observational noise (see Section \ref{Section:Noise power spectrum}).

\subsection{Power spectrum of residual surface-brightness fluctuations} 
\label{Section:HST_power_spectrum_residuals}

To quantify the overall residual surface-brightness fluctuations, we follow the procedure introduced in Paper I and compute the azimuthally-averaged power spectrum of the residual image within the mask covering the lensed images (shown in Fig.~\ref{fig:Galfit_J0252}). To achieve this, we set the flux values of the pixels located outside the mask to zero and calculate the two-dimensional discrete Fourier transform (DFT) of the masked residual image, using the Python package \textsc{numpy.fft}\footnote{https://docs.scipy.org/doc/numpy/reference/routines.fft.html}. The squared magnitude of the Fourier coefficient assigned to each pixel yields the two-dimensional power spectrum. Assuming isotropy of the modelled potential perturbations $\delta\psi(\textbf{\textit{x}})$, we average the power-spectrum values along a set of ten equidistant concentric annuli spanning from $k_{\rm{min}}= $ 0.88 to $k_{\rm{max}}= $ 16.79 $\mathrm{arcsec^{-1}}$ (corresponding to the spatial scales between $\lambda_{\rm{min}}= $ 0.22 and $\lambda_{\rm{max}}= $ 4.65 kpc at the redshift of the lens galaxy $z_{\rm l} = 0.280$).

\subsection{Noise correction} 
\label{Section:Noise power spectrum}

In order to characterise the noise properties in the analysed HST-image, we create a sample of 20 selected blank-sky regions with the same size (121 by 121 pixels), located in proximity to the lens system. The first rough estimate of the noise level is given by the standard deviation of the flux values in this blank-sky sample: $\sigma_{\rm sky}= 0.002 \ \mathrm{e^- \ sec^{-1}}$. However, this estimate does not take into account the photon-shot (Poisson-distributed) noise, which depends on the number of detected photons and, consequently, varies from pixel to pixel. 

A more precise description of noise properties, including both the sky-background and the photon-shot noise, is provided by the noise-sigma map, which quantifies the standard deviation of noise for each pixel separately. Considering that the Poisson variance of the photon counts is similar to the measured number of photons and the raw HST images of the lens are drizzle-combined using an inverse-variance map weighting, we construct the noise-sigma map for our drizzled HST science image according to the following formula: 
\begin{equation}
\sigma_{n} = \sqrt{N/W + \sigma_{\rm sky}^2}, 
\label{eq:noise_sigma_map}
\end{equation}
where $N$ is the number of photo-electrons per second detected in a particular pixel (after the sky-background subtraction) and $W$ is the weight of this pixel taken from the weight map of our image provided by the drizzling pipeline. This noise-sigma map is also used in the Bayesian smooth-lens modelling procedure, presented in Section \ref{Section:Lens_Model_J0252}. Since the Poisson noise approaches Gaussian noise for large number counts, as is the case for the studied image, in the remaining part of our analysis we approximate the photon-shot noise by an additive Gaussian noise $\mathcal{N}(0, \sigma_{n}^2)$ with a variance $\sigma_{n}^2$ adapted to the flux value in a particular pixel.

However, due to noise correlations in the drizzled images and the charge-transfer inefficiency (CTI), discussed in more detail in Paper I, the noise correction of the residual surface-brightness fluctuations requires a more extended approach. To mimic the true noise properties in the science image, we use the selected blank-sky regions to generate a sample of 20 scaled sky-background realisations, which account for both the realistic noise-correlation pattern and the spatially-varying flux-dependent photon-shot noise. This is implemented by first dividing the flux values of the blank-sky regions by their standard deviation and subsequently multiplying them by the noise-sigma map of the science image (equation \ref{eq:noise_sigma_map}). For consistency reasons, these scaled sky-background realisations are finally overlaid with the same mask as the one used in the analysis of the real data (Section \ref{Section:HST_power_spectrum_J0252}). The average power spectrum measured in this sample is used as our best estimate for the total noise power spectrum of the science image (see Paper I for a more thorough discussion). 

\begin{figure}
 \includegraphics[width=\columnwidth]{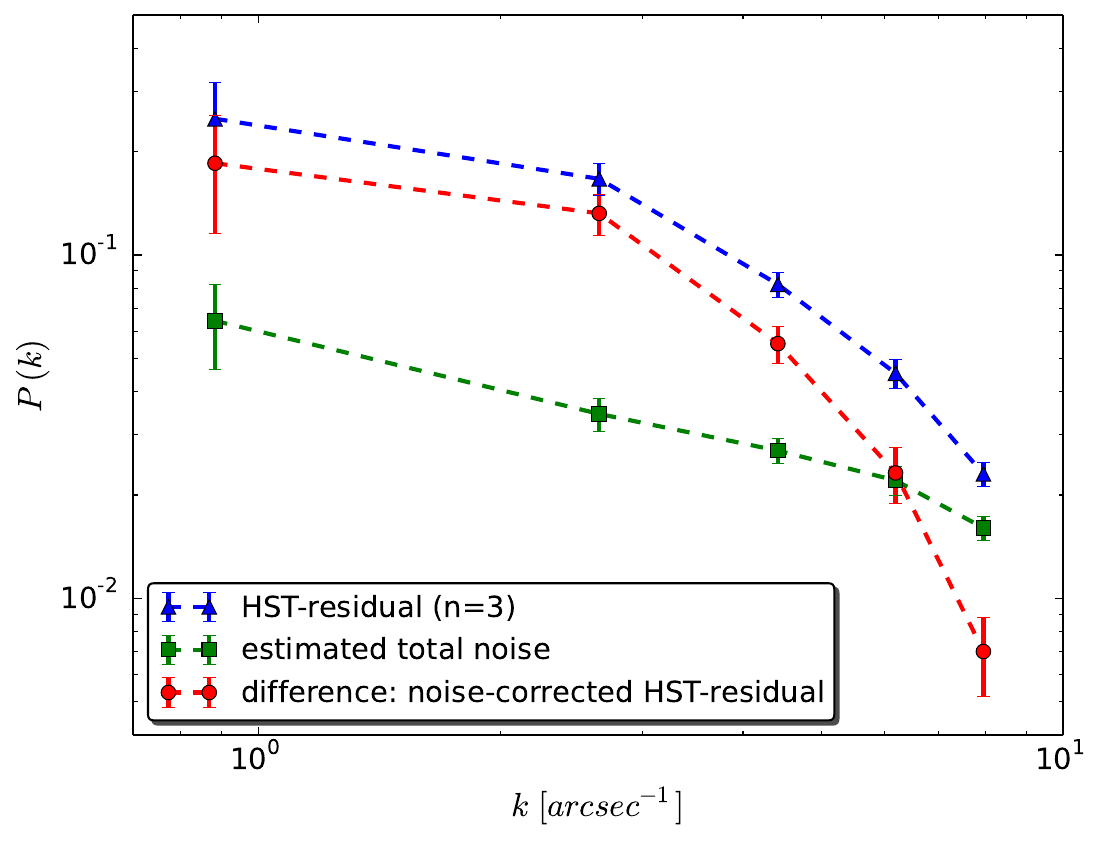}
 \caption{Power spectrum of residual surface-brightness fluctuations remaining in the lensed images of SDSS J0252+0039 after the lens-galaxy subtraction, the smooth lens modelling and the noise correction. The difference between the power spectrum of residual surface-brightness fluctuations after the smooth lens modelling with $n = 3$ \textit{(blue line)} and the estimated total noise power spectrum \textit{(green line)} constitutes our upper limit on the power spectrum of surface-brightness anomalies due to mass structure in the lens galaxy SDSS J0252+0039 \textit{(red line)}.} 
 \label{fig:power_spectra}
\end{figure}

Finally, we use this estimated total noise power spectrum to perform the noise correction of the residual surface-brightness fluctuations. We assume that the observational noise and the potential fluctuations $\delta\psi$, perturbing the smooth lensing potential, are independent stochastic processes and consider the corresponding power spectra to be additive. This allows us to subtract the estimated noise power spectrum from the power spectrum of the total residual surface-brightness fluctuations, as illustrated in Fig.~\ref{fig:power_spectra}. The difference of these two power spectra is treated in our further analysis as an upper limit to the power spectrum of the surface-brightness anomalies $P_{\delta I}(k)$ caused by the hypothetical small-scale mass structures in the lens galaxy. 

However, a comparison between the power spectrum of the revealed residual surface-brightness fluctuations with the estimated noise power spectrum shows that for the highest considered $k$-values (corresponding to scales below three pixels) the residual fluctuations reach the noise level. This indicates that no surface-brightness anomalies are detected on these scales. For this reason, in our further analysis we consider only the perturbation wave numbers ranging from $k_{\rm{min}}= $ 0.88 to $k_{\rm{max}}= $ 7.95 $\mathrm{arcsec^{-1}}$, which corresponds to the spatial scales between $\lambda_{\rm{min}}= $ 0.52 and $\lambda_{\rm{max}}= $ 4.65 kpc at the redshift of the lens galaxy, see Fig.~\ref{fig:power_spectra}. Performing the analysis on scales above three pixels together with our choice of $n=3$ in the smooth-lens-modelling procedure allows us to neglect the effects of small errors in the PSF (with $\mathrm{FWHM} = 0.07 \ \mathrm{arcsec})$, the correlations between adjacent pixels due to drizzling and the possible residual errors in the source-light modelling.

\section{Catalogue of perturbed lensed images}
\label{Section:mock_perturbed_lenses_J0252}

In order to interpret the power spectrum of the surface-brightness anomalies measured in the observed image of SDSS J0252+0039, in this section we generate a catalogue of simulated lensed images which mimic these observations but are perturbed by Gaussian potential perturbations $\delta\psi_{\rm{GRF}}(\textbf{\textit{x}})$ from the power-law power spectrum $P_{\delta\psi}(k;\sigma^2_{\delta\psi}, \beta)$ (see Section \ref{Section:GRF}) on a grid containing $100 \times 100$ different combinations of $\beta$ and $\sigma^2_{\delta\psi}$. The considered values of $\beta$ are equidistant within the interval $[3, 8]$, whereas $\sigma^2_{\delta\psi}$ (within the studied field-of-view) is varied by obtaining $100$ values evenly spaced in the logarithmic range $[10^{-6}, 10^{-1}]$.   

\begin{figure*}
\includegraphics{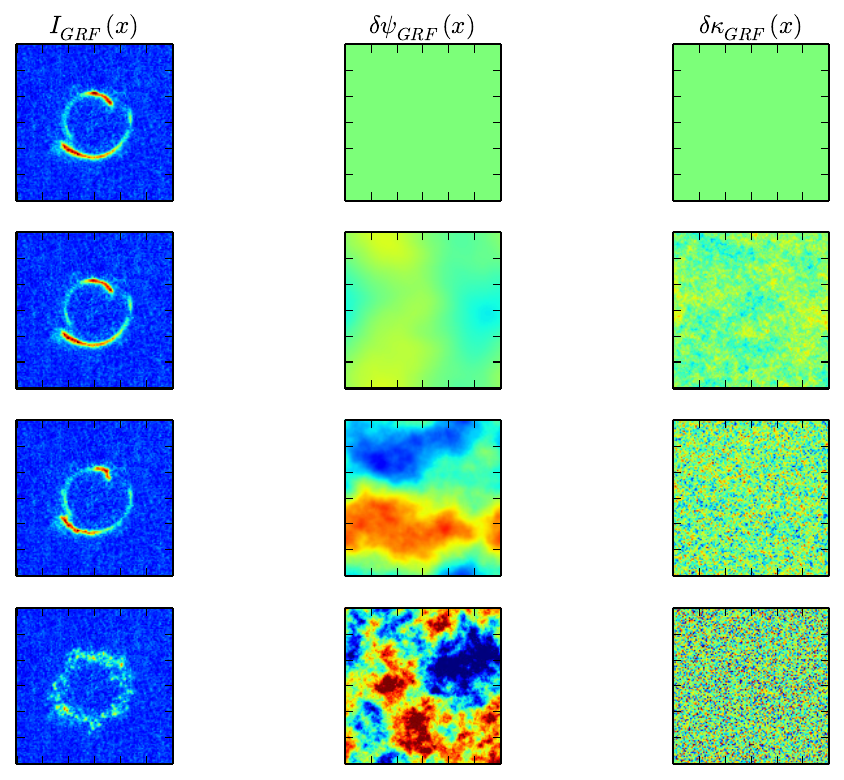}
\caption{Mock surface-brightness anomalies induced in the gravitational arcs of SDSS J0252+0039 by Gaussian potential perturbations $\delta\psi_{\rm{GRF}}(\textbf{\textit{x}})$ in the lens galaxy, fully characterised by the power spectrum $P_{\delta\psi}(k) \propto \sigma^2_{\delta\psi} \times k^{-\beta}$. \textit{Left column}: unperturbed lensed images, i.e. the reconstructed source galaxy lensed through the best-fitting smooth-lens model \textit{(upper panel)} and three examples of perturbed lensed images (for $\sigma^2_{\delta\psi} = 2.154 \times 10^{-5}$ and $\beta = 5.5 $, $\sigma^2_{\delta\psi} = 2.783 \times 10^{-4}$ and $\beta = 4.25 $, $\sigma^2_{\delta\psi} = 10^{-3}$ and $\beta = 3 $; \textit{from top to bottom}) with a realistic noise realisation overlaid for visualisation purposes. \textit{Middle column}: the underlying realisations of $\delta\psi_{\rm{GRF}}(\textbf{\textit{x}})$. \textit{Right column}: the corresponding convergence (i.e. surface mass density) perturbations $\delta\kappa_{\rm{GRF}}(\textbf{\textit{x}})=\frac{1}{2}\nabla^2 \delta \psi_{\rm{GRF}}(\textbf{\textit{x}})$.}
\label{fig:mock_perturbed_noise}
\end{figure*}

For each considered combination of $\beta$ and $\sigma^2_{\delta\psi}$, we generate a pixelated realisation of $\delta\psi_{\rm{GRF}}(\textbf{\textit{x}})$ from the corresponding power spectrum $P_{\delta\psi}(k;\sigma^2_{\delta\psi}, \beta)$ (equations \ref{eq:power_law_PS_J0252} and \ref{eq:A_J0252}) and superimpose it on the best-fitting smooth lensing potential $\psi_0(\textbf{\textit{x}})$ of SDSS J0252+0039, obtained for the observational data in Section \ref{Section:Lens_Model_J0252}. We then apply this perturbed lensing potential $\psi_0(\textbf{\textit{x}}) + \delta\psi_{\rm{GRF}}(\textbf{\textit{x}})$ to repeat the lensing operation of the previously reconstructed pixellated surface-brightness distribution of the background source galaxy and finally obtain the resulting perturbed lensed image. Fig.~\ref{fig:mock_perturbed_noise} shows three examples of these mock images for different values of $\sigma^2_{\delta\psi}$ and $\beta$, together with the underlying realisations of $\delta\psi_{\rm{GRF}}(\textbf{\textit{x}})$ and $\delta\kappa_{\rm{GRF}}(\textbf{\textit{x}})$.

\begin{figure}
 \includegraphics[width=\columnwidth]{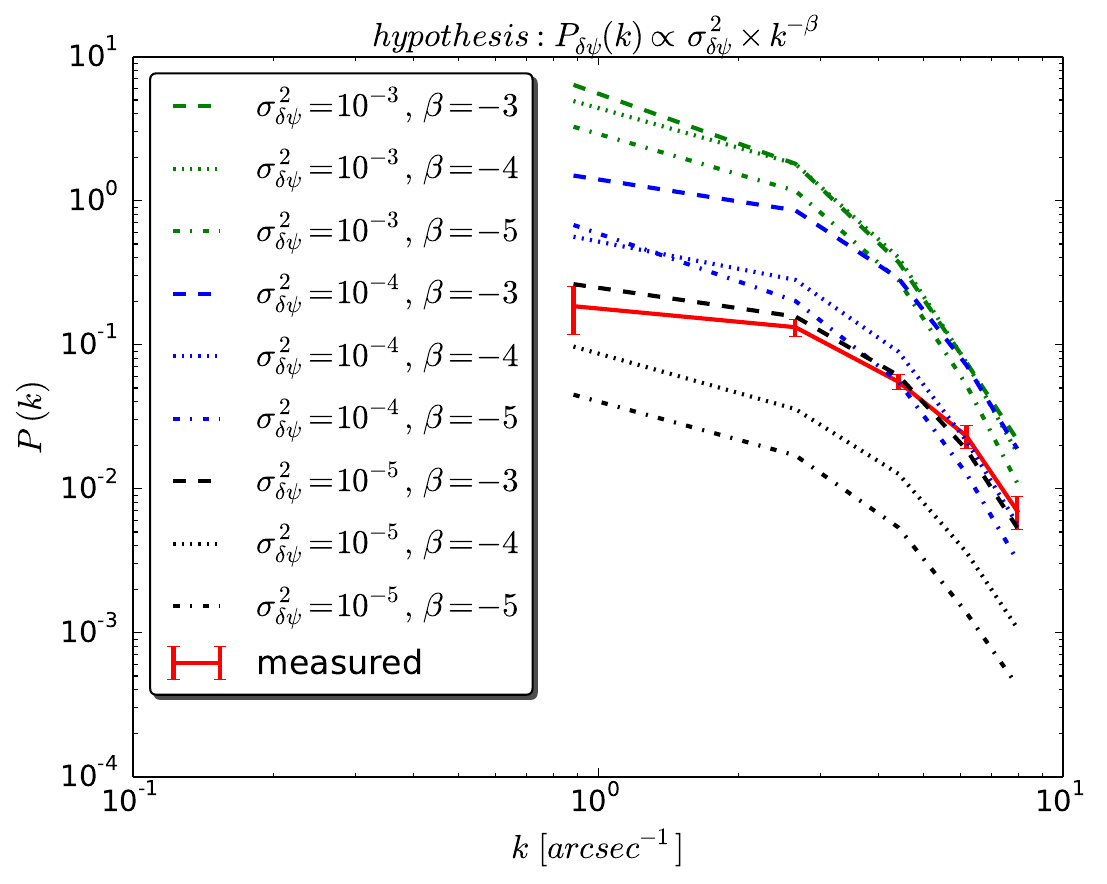}
 \caption{Upper limit on the power spectrum of surface-brightness anomalies due to mass structure in the lens galaxy SDSS J0252+0039 \textit{(red line)}, corresponding to the red line in Fig.~\ref{fig:power_spectra}, in comparison to a mock catalogue containing power spectra of surface-brightness anomalies induced by Gaussian potential perturbations with known values of the integrated variance $\sigma^2_{\delta\psi}$ and the power-law power-spectrum slope $\beta$ according to equations \ref{eq:power_law_PS_J0252} and \ref{eq:A_J0252} \textit{(dashed lines)}. For clarity of presentation, we show only 9 out of $10^4$ mock power spectra in our catalogue.}
 \label{fig:mockPS}
\end{figure}

To obtain the power spectra of surface-brightness anomalies for these perturbed mock lensed images, we follow the methodology applied to the observational data, i.e. we subtract the best-fitting smooth-lens model from each of the mocks and compute the power spectrum of the residual surface-brightness fluctuations within the same mask and using the same set of ten bins for the azimuthal averaging as for the observational data. We stress that in this paper, for simplicity, we do not model the perturbed lensed images individually, but assume that the same best-fitting smooth lens model that was inferred for the data holds for all the perturbed mocks. We plan to investigate the validity of this assumption in our future paper (Bayer et al. in prep). Moreover, to reduce the sample variance, we consider ten different realisations of $\delta\psi_{\rm{GRF}}(\textbf{\textit{x}})$ for each combination of $\sigma^2_{\delta\psi}$ and $\beta$ and average the resulting power spectra over this sample. Fig.~\ref{fig:mockPS} presents a few examples of the averaged mock power spectra. A comparison of these to the real measurement, illustrated in Fig.~\ref{fig:mockPS}, allows us to infer exclusion probabilities for the considered portion of the parameter space spanned by $\sigma^2_{\delta\psi}$ and $\beta$, which is discussed in the next section.

\section{Constraints on the sub-galactic matter power spectrum} 
\label{Section:Results}

In this section, we first discuss our approach to assess whether a particular matter-power-spectrum model (i.e. combination of $\sigma^2_{\delta\psi}$ and $\beta$) leads to surface-brightness anomalies that are significantly different from the measured residual surface-brightness fluctuations. Subsequently, we present and discuss the resulting constraints on the statistical properties of the potential perturbations in the lens galaxy SDSS J0252+0039 (Section \ref{Section:Exclusion_Probablility}). Finally, we express these constraints in terms of the variance of the resulting deflection-angle perturbations (Section \ref{Section:Results_deflection}), the dimensionless convergence power spectrum $\Delta^{2}_{\delta\kappa}(k)$ and the aperture mass on three specific spatial scales: 0.5, 1 and 3 kpc (Section \ref{Section:Results_convergence}).

\subsection{Exclusion probability of matter-power-spectrum models}
\label{Section:Exclusion_Probablility}

A comparison between the power spectra of the mock surface-brightness anomalies and the power spectrum of the actually observed residual surface-brightness fluctuations enables us to determine exclusion probabilities for different combinations of $\sigma^2_{\delta\psi}$ and $\beta$. As discussed in Section \ref{Section:Analysis overview}, we conservatively treat the estimated residual surface-brightness fluctuations as an upper limit to the anomalies caused solely by potential perturbations in the lens galaxy. Consequently, we rule out a particular matter power-spectrum model only if the power spectrum of the resulting surface-brightness anomalies exceeds the measured power spectrum on all considered scales, i.e. in all five analysed $k$-bins between $k_{\rm{min}}= $ 0.88 and $k_{\rm{max}}= $ 7.95 $\mathrm{arcsec^{-1}}$. 

In order to determine the exclusion probability, we treat the power-spectrum value measured in the data for each individual $k$-bin as the expected value of a Gaussian random variable $P^{D}_{\delta I}(k)$ with the standard deviation (corresponding to the error bars in Fig.~\ref{fig:mockPS}) estimated based on the variance in our sample of mock noise realisations introduced in Section~\ref{Section:Noise power spectrum}. Under this assumption, we first compute the probability that the power spectrum of the mock surface-brightness anomalies for a given matter-power-spectrum model exceeds the actually measured power spectrum for each $k$-bin separately in the following way:
\begin{equation}
\begin{split}
& P_{\rm excl} \Big( \sigma^2_{\delta\psi},\beta,k \Big) = \\
& \text{Prob} \left( \,P^{D}_{\delta I}(k)<P^{M}_{\delta I}(k) \mid \sigma^2_{\delta\psi},\ \beta\ , \ P_{\delta\psi}(k) \propto \sigma^2_{\delta\psi} \times k^{-\beta}\right), 
\end{split}
\end{equation}
with $P^{M}_{\delta I}(k)$  denoting the mock power spectrum for a particular choice of $ \sigma^2_{\delta\psi}$ and $\beta$. Assuming that our exclusion-probability estimates for different perturbation wavenumbers $k$ are statistically independent, the final exclusion probability for a particular matter-power-spectrum model can be calculated as the product of the exclusion probabilities in all considered $k$-bins:
\begin{equation}
P_{\rm excl} \Big( \sigma^2_{\delta\psi},\beta \Big)= \prod_{k}P_{\rm excl} \Big( \sigma^2_{\delta\psi},\beta,k \Big).
\label{Eq.Exclusion}
\end{equation}

\begin{figure*}
\centering 
\includegraphics{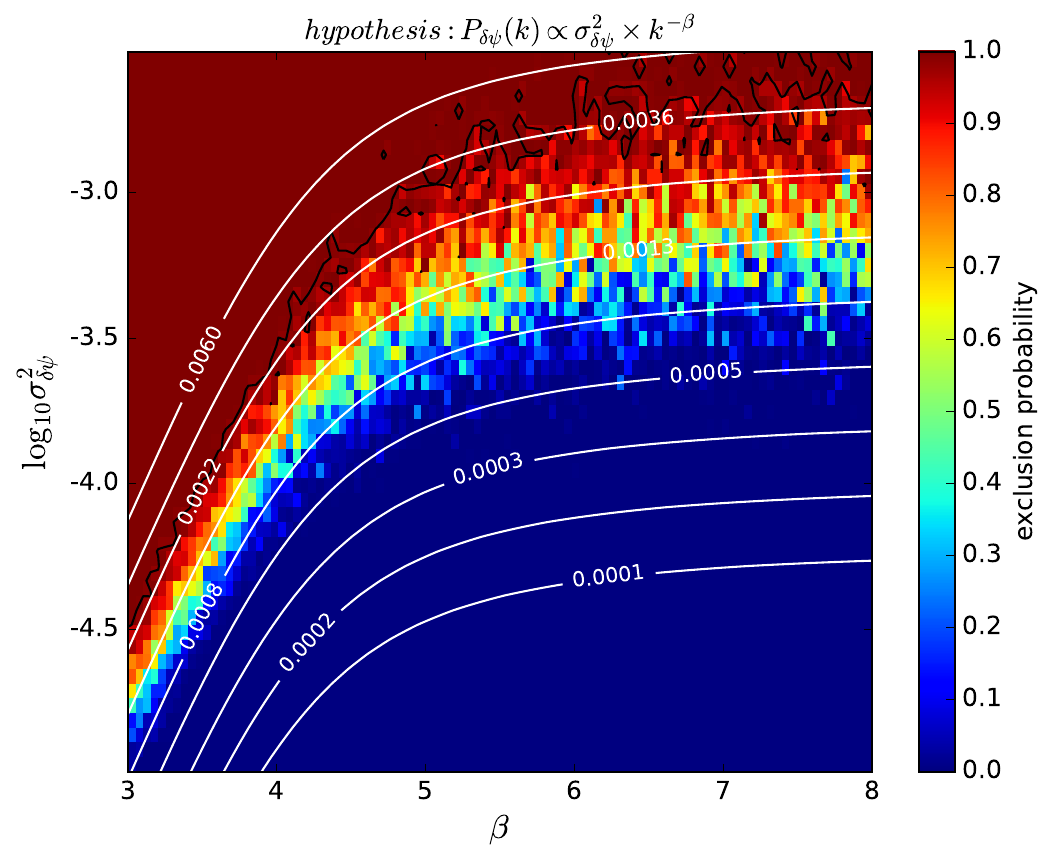}
\caption{Exclusion probabilities $P_{\rm excl} \Big( \sigma^2_{\delta\psi},\beta \Big)$ for a subset of considered sub-galactic matter-power-spectrum models, inferred from our analysis of the lens system SDSS J0252+0039. Each point of this exclusion plot corresponds to a particular combination of the integrated variance $\sigma^2_{\delta\psi}$ and the power-law slope $\beta$, assumed to fully characterize the power spectrum of the hypothetical GRF potential perturbations $P_{\delta\psi}(k) \propto \sigma^2_{\delta\psi} \times k^{-\beta}$ in the lens galaxy. The superposed \textit{black isoprobability line} indicates models with exclusion probability larger than 99 per cent. The \textit{white contour lines} connect matter-power-spectrum models with the same variance of the associated deflection-angle perturbations $\sigma^2_{\delta\alpha}$ in the analysed field-of-view.}
\label{fig:exclusion_plot_extended_cut}
\end{figure*}

The inferred exclusion probabilities for the considered range of matter-power-spectrum models (i.e. combinations of $\sigma^2_{\delta\psi}$ and $\beta$) are presented in Fig.~\ref{fig:exclusion_plot_extended_cut}. The depicted grid contains $50$ different values for $\sigma^2_{\delta\psi}$, evenly spaced in the logarithmic range $[10^{-5.0},10^{-2.5}]$, and $100$ values for $\beta$, equidistant in the interval $[3, 8]$. Colour-coded is the inferred exclusion probability of the corresponding matter-power-spectrum models $P_{\rm excl} \Big( \sigma^2_{\delta\psi},\beta \Big)$, defined as the probability that the resulting surface-brightness anomalies exceed the observed ones on all considered spatial scales. The superposed (black) isoprobability line indicates models with exclusion probability larger than 99 per cent.

Based on the exclusion plot presented in Fig.~\ref{fig:exclusion_plot_extended_cut}, we rule out matter-power-spectrum models with $\sigma^2_{\delta\psi}$
exceeding $\sim 10^{-2.5}$ (on the spatial scales between $L^{-1}$ and pixel scale$^{-1}$ and within the range of the power-law power-spectrum slope $3\leq\beta\leq8$) at the 99 per cent confidence level. This corresponds to the upper bound on the standard deviation of such potential fluctuations $\sigma_{\delta\psi} \leq 0.06$. The excluded models would lead to power spectra of surface-brightness anomalies that significantly exceed the observed upper limits on all considered scales. 
Moreover, we find that for a specific value of the integrated variance $\sigma^2_{\delta\psi}$ the exclusion probability depends on the power-spectrum slope $\beta$ and, thus, on the exact distribution of the integrated variance between the different spatial scales. Shallower slopes, assigning a larger fraction of $\sigma^2_{\delta\psi}$ to small scales, i.e. high $k$-modes, are found more likely to be excluded. This leads to the conclusion that potential fluctuations on smaller spatial scales have a stronger perturbative effect on the lensed images than potential fluctuations on larger scales.

\subsection{Upper limits on the deflection-angle perturbations} 
\label{Section:Results_deflection}

To explain the asymptotic shape of the exclusion plot, presented in Fig.~\ref{fig:exclusion_plot_extended_cut}, we convert our original constraints on the power spectrum of the potential perturbations $P_{\delta\psi}\Big(k, \sigma^2_{\delta\psi},\beta \Big)$ into constraints on the power spectrum of the corresponding perturbations in the deflection angle $P_{\delta\alpha} \Big(k, \sigma^2_{\delta\psi},\beta \Big)$, making use of the following relation: 
\begin{equation}
P_{\delta\alpha}(k)= 4\pi^{2} k^{2}P_{\delta\psi}(k).
\label{PS_delta_alpha_J0252}
\end{equation} 
According to equation (\ref{PS_delta_alpha_J0252}), the slope of $P_{\delta\alpha} \Big(k, \sigma^2_{\delta\psi},\beta \Big)$ decreases by 2.0 (i.e. becomes shallower) with respect to the slope of $P_{\delta\psi}\Big(k, \sigma^2_{\delta\psi},\beta \Big)$. The total variance in the differential deflection angle $\sigma^2_{\delta\alpha}$ over the analysed field-of-view is obtained by integrating $P_{\delta\alpha}(k)$ over all pixels of the two-dimensional Fourier grid:
\begin{equation}
\sigma^2_{\delta\alpha}\Big( \sigma^2_{\delta\psi},\beta \Big) = \int_{k_{x}}\int_{k_{y}} P_{\delta\alpha} \Big( \sqrt{k_{x}^2+k_{y}^2}, \sigma^2_{\delta\psi},\beta \Big) \ dk_{x} \ dk_{y}.
\label{eq:total_variance_alpha}
\end{equation}

The result is presented in Fig.~\ref{fig:exclusion_plot_extended_cut}, where the overlaid (white) isocontours correspond to the same values of $\sigma^2_{\delta\alpha}$. These isocontours almost perfectly follow the overall shape of the exclusion limits, which indicates a strong correlation between the exclusion probability of a matter-power-spectrum model and the total variance of the associated deflection-angle perturbations. This suggests that $\sigma^2_{\delta\alpha}$ is the fundamental quantity that determines the level of the resulting surface-brightness anomalies in the lensed images. Consequently, the exclusion probability is almost insensitive to the slope $\beta$ of the $P_{\delta\alpha}(k)$, i.e. the distribution of the total variance in the deflection angle over the different length scales. Based on Fig.~\ref{fig:exclusion_plot_extended_cut}, we exclude (with 99 per cent probability) a matter-power-spectrum model if the corresponding total variance in the differential deflection field is larger than $6\times10^{-3}$, independently of the slope $\beta$. This insight is valuable for a future analysis of additional lens systems, which instead of considering potential perturbations $\delta\psi_{\rm{GRF}}(\textbf{\textit{x}})$ might more efficiently focus on the corresponding deflection-angle perturbations  $\bm{\delta\alpha}_{\rm{GRF}}(\textbf{\textit{x}})$. The inferred threshold value of $\sigma^2_{\delta\alpha}$, however, might vary between different lens systems and depend on the chosen field-of-view, the PSF and the signal-to-noise ratio of the analysed image.

For completeness, we derive the corresponding constraints on the dimensionless deflection-power spectrum defined as follows:
\begin{equation}
\Delta^{2}_{\delta\alpha}(k)\equiv 2\pi k^{2} P_{\delta\alpha}(k)
\label{delta_alpha_J0252}
\end{equation}
for the spatial scales of 0.5, 1 and 3 kpc, see Fig.~\ref{fig:Exclusion_panel_combined}. These exclude $\Delta^{2}_{\delta\alpha}(k)$ larger than $0.001$ on all considered spatial scales at the 99 per cent confidence level.

\subsection{Upper limits on the convergence power spectrum} 
\label{Section:Results_convergence}

Finally, the inferred constraints on the power spectrum of the potential perturbations $P_{\delta\psi}(k)$ can be translated into constraints on the associated power spectrum of the convergence (i.e. surface mass density) perturbations $P_{\delta\kappa}(k)$. By expressing equation (\ref{Poisson_delta_J0252}) in Fourier space, $P_{\delta\kappa}(k)$ can be related to $P_{\delta\psi}(k)$ as follows: 
\begin{equation}
P_{\delta\kappa}(k)=4\pi^{4}k^{4}P_{\delta\psi}(k).
\end{equation}
For a future comparison with the $\Lambda$CDM predictions, we express these constraints in terms of the dimensionless convergence power spectrum:
\begin{equation}
\Delta^{2}_{\delta\kappa}(k)\equiv2\pi k^{2} P_{\delta\kappa}(k),
\label{delta_kappa_J0252}
\end{equation}
which quantifies the contribution of a particular length scale $\lambda=k^{-1}$ to the total variance of the convergence perturbations. 

\begin{figure*}
\centering
\includegraphics[scale=0.93]
{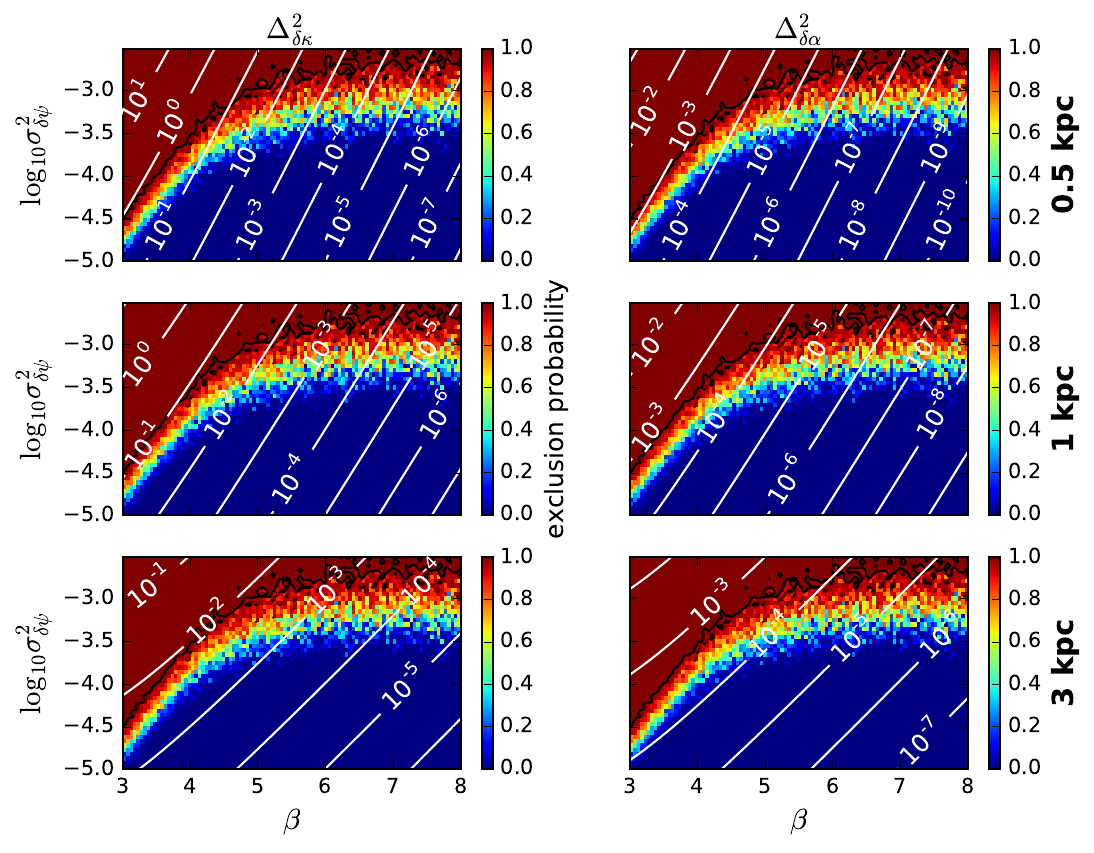}
\caption{Upper-limit constraints on the dimensionless convergence power spectrum $\Delta^{2}_{\delta\kappa}$ and the dimensionless differential-deflection power spectrum $\Delta^{2}_{\delta\alpha}$ in the lens galaxy SDSS J0252+0039 for three different sub-galactic scales. Each point of the exclusion plot corresponds to a particular sub-galactic matter-power-spectrum model $P_{\delta\psi}(k) \propto \sigma^2_{\delta\psi} \times k^{-\beta}$, as specified in Fig.~\ref{fig:exclusion_plot_extended_cut}. The \textit{white contour lines} connect models with the same value of $\Delta^{2}_{\delta\kappa}$ \textit{(left panel)} and $\Delta^{2}_{\delta\alpha}$ \textit{(right panel)} on a particular scale. Overlaid in black is the 0.99-contour of the exclusion probability. Note that $\beta$ represents the power-spectrum slope of the originally investigated potential perturbations $\delta\psi_{\rm{GRF}}(\textbf{\textit{x}})$. The slopes of the corresponding power spectra of the convergence perturbations $\delta\kappa_{\rm{GRF}}(\textbf{\textit{x}})$ and the deflection perturbations $\delta\alpha_{\rm{GRF}}(\textbf{\textit{x}})$ decrease by 4 and 2, respectively (in terms of the absolute value).}
\label{fig:Exclusion_panel_combined}
\end{figure*}

We determine $\Delta^{2}_{\delta\kappa}(k)$ corresponding to each combination of $\sigma^2_{\delta\psi}$ and $\beta$ for three different spatial scales: 0.5, 1 and 3 kpc. The smallest scale of 0.5 kpc corresponds to $\sim$3 pixels, which is the smallest spatial scale considered in our analysis. The largest considered scale, on the other hand, is limited by the size of the lensed images (gravitational arcs) in SDSS J0252+0039. ~Fig.~\ref{fig:Exclusion_panel_combined} shows the resulting contour lines connecting matter-power-spectrum models with the same value of $\Delta^{2}_{\delta\kappa}$ inferred on a particular scale, overlaid on the original exclusion plot. Based on Fig.~\ref{fig:Exclusion_panel_combined}, we rule out matter-power-spectrum models with $\Delta^{2}_{\delta\kappa}$ larger than $1$ on 0.5-kpc scale, larger than $0.1$ on 1-kpc scale and larger than $0.01$ on 3-kpc scale, at the 99 per cent confidence level.

These results can be additionally interpreted in terms of the standard deviation of the total convergence perturbation $\sigma_{\delta\kappa}(\lambda) \equiv\sqrt{\Delta^{2}_{\delta\kappa}(\lambda)}$ within the aperture diameter equal to the considered scale $\lambda$ in an infinitely large sample of circular regions, randomly chosen in proximity to the Einstein radius. For the lensing-mass distribution of SDSS J0252+0039 we infer the following upper limits on this standard deviation (on the spatial scales between $L^{-1}$ and pixel scale$^{-1}$, within the range of the power-law power-spectrum slope $3\leq\beta\leq8$): $\sigma_{\delta\kappa}(0.5 \  \mathrm{kpc})<1$ on 0.5-kpc scale, $\sigma_{\delta\kappa}(1 \  \mathrm{kpc})<0.3$ on 1-kpc scale and $\sigma_{\delta\kappa}(3 \ \mathrm{kpc})<0.1$ on 3-kpc scale. With the critical surface-mass density for SDSS J0252+0039 $\Sigma_{\rm{cr}} \approx 4 \times 10^8 \ M_\odot \  \mathrm{kpc^{-2}}$, these constraints can be translated into upper limits on the integrated standard deviation $\sigma_{AM}(\lambda)$ in the aperture mass (within an aperture of diameter $\lambda$ in the lens plane) in the inner region of the lens galaxy SDSS J0252+0039: $\sigma_{AM}(0.5 \  \mathrm{kpc})< 0.8 \times 10^8 M_\odot$, $\sigma_{AM}(1 \  \mathrm{kpc})< 1 \times 10^8 M_\odot$ and $\sigma_{AM}(3 \ \mathrm{kpc})< 3 \times 10^8 M_\odot$, at the 99 per cent confidence level.

\section{Discussion} 
\label{Section:Discussion}

In order to compare the derived upper-limit constraints with the predictions from the $\Lambda$CDM model, we now provide a simple heuristic estimation of the expected dimensionless convergence power spectrum $\Delta^{2}_{\delta\kappa}(\lambda)$ due to CDM subhaloes in the dark-matter halo of SDSS J0252+0039. For the sake of simplicity, in our estimation we only consider the contribution from CDM subhaloes, neglecting both possible line-of-sight haloes and any other fluctuations in the baryonic or dark-matter distribution of the lens galaxy. Furthermore, we assume the projected substructure mass fraction of 0.005 near the Einstein radius of the lens halo in the subhalo mass range between $4 \times 10^6$ and $4 \times 10^9 M_\odot$ \citep{Vegetti_statistics_2009}. Finally, we conservatively treat the subhaloes as point masses and apply the Poisson statistics to analytically predict their abundance from the CDM substructure mass function of the form $dN/dM \propto M^{-1.9}$ \citep{Springel_Aquarium}, thus neglecting the possible suppression of the substructure population due to baryonic processes \citep{Despali_LOS}. Taking into consideration that the studied lens galaxy is well described by the Singular Isothermal Sphere model and, thus, its convergence $\kappa \sim 0.5$ in proximity to the Einstein radius, we estimate the following upper limits on $\Delta^{2}_{\delta\kappa}(\lambda)$ due to CDM subhaloes : $\Delta^{2}_{\rm{CDM}}(0.5 \  \mathrm{kpc})<10^{-3}$, $\Delta^{2}_{\rm{CDM}}(1 \  \mathrm{kpc})<4 \times 10^{-4}$ and $\Delta^{2}_{\rm{CDM}}(3 \  \mathrm{kpc})<10^{-4}$. While this point-mass approach is solely an approximation, it provides a conservative upper limit on the convergence contribution from CDM subhaloes. Any other choice of the subhalo density profile, for example the more realistic Navarro-Frenk-White (NFW) density profile, would only lower the predicted $\Delta^{2}_{\rm{CDM}}(\lambda)$.

A comparison of this estimation with the inferred observational constraints on $\Delta^{2}_{\delta\kappa}$ for a flat convergence power spectrum (corresponding to the power-spectrum slope $\beta=4$ for the potential perturbations; see Fig.~\ref{fig:Exclusion_panel_combined}) leads to the conclusion that the estimated contribution from CDM subhaloes lies significantly below the observational upper limits on all considered scales. This preliminary conclusion does not change even if we take into account that in reality the total number of haloes perturbing the lensed images might be a few times higher than the estimated number of subhaloes, due to the possible presence of unbound haloes along the line-of-sight \citep{Despali_LOS}. 

We attribute the above discrepancy mainly to the fact that, unlike our heuristic $\Lambda$CDM-based predictions, the inferred observational upper-limit constraints refer to the total (dark and baryonic) mass distribution projected along the line-of-sight, including the complex baryonic and dark matter structure of the lens galaxy \citep[e.g.][]{Gilman, Hsueh_disk} and the line-of-sight haloes. Further research is required to adequately compare these observational limits with hydrodynamical simulations, which model not only the formation of dark-matter haloes and subhaloes, but also include baryonic processes and their effect on the overall mass distribution in galaxies.

\section{Summary and conclusions}
\label{Section:Conclusions}

The alternative dark-matter models and galaxy-formation scenarios predict significantly different levels of mass structure on the sub-galactic scales \citep[e.g.][]{Lovell2014}. In this work, we have introduced a novel methodology to observationally constrain the statistical properties of such small-scale density fluctuations in the total projected mass distribution of massive elliptical lens galaxies by means of the power-spectrum analysis of surface-brightness anomalies measured in highly-magnified galaxy-scale Einstein rings and gravitational arcs. Our approach is based on the theoretical framework introduced by \cite{Saikat}. 

The pilot application of the presented methodology to the lens system SDSS J0252+0039 from the SLACS Survey leads to the following conclusions:  
\begin{enumerate}

\item
The enhanced intrinsic source-galaxy structure in the analysed U-band data requires a higher source-grid resolution and leads to more severe degeneracies between the source and lens models than it was the case for the I-band data previously modelled by \cite{Vegetti2014}. Whereas this degeneracy is less problematic when trying to identify individual subhaloes with masses above the detection limit, see \cite{Vegetti2014}, alleviating it is crucial when performing a power-spectrum analysis. In this paper, we have addressed this degeneracy by lowering the resolution of the source reconstruction in the smooth lens modelling to suppress the absorption of the potential perturbations and the observational noise into the source structure. This strategy has been shown to be effective in the performance test of our methodology discussed in Paper I.

\vspace{0.1cm}

\item 
Our analysis of SDSS J0252+0039 rules out the presence of Gaussian potential perturbations $\delta\psi_{\rm{GRF}}(\textbf{\textit{x}})$ with the variance $\sigma^2_{\delta\psi}$ exceeding $\sim 10^{-2.5}$  at 99\% C.L. (on the spatial scales between $L^{-1}= 0.88 \ \mathrm{arcsec^{-1}}$ and pixel scale$^{-1}=16.79 \ \mathrm{arcsec^{-1}}$ and within the range of the GRF power-spectrum slope $3\leq\beta\leq8$).
\vspace{0.1cm}

\item 
In order to account for the effect of the chosen field-of-view, we infer the corresponding constraints on the dimensionless convergence power spectrum $\Delta^{2}_{\delta\kappa}(\lambda)$ on three different sub-galactic scales and rule out matter-power-spectrum models with $\Delta^{2}_{\delta\kappa}(0.5 \  \mathrm{kpc})>1$ on 0.5-kpc scale, $\Delta^{2}_{\delta\kappa}(1 \  \mathrm{kpc})>0.1$ on 1-kpc scale and $\Delta^{2}_{\delta\kappa}(3 \  \mathrm{kpc})>0.01$ on 3-kpc scale (at the 99\% C.L.). 
\vspace{0.1cm}

\item
The inferred constraints can be translated into to the following limits on the standard deviation $\sigma_{\rm AM}(\lambda)$ of the aperture mass (integrated within a cylinder with the respective diameter $\lambda$ in the lens plane) in proximity to the Einstein radius of the lens galaxy: $\sigma_{\rm AM}(0.5 \  \mathrm{kpc})< 0.8 \times 10^8 M_\odot$, $\sigma_{\rm AM}(1 \  \mathrm{kpc})< 1 \times 10^8 M_\odot$ and $\sigma_{\rm AM}(3 \ \mathrm{kpc})< 3 \times 10^8 M_\odot$ (at the 99\% C.L.). 

\vspace{0.1cm}

\item 
We find that the fundamental quantity that determines the level of surface-brightness anomalies in the lensed images and, thus, the probability of the matter-power-spectrum model exclusion, is the total variance in the differential deflection angle $\sigma^2_{\delta\alpha}$ (on the spatial scales between $L^{-1}$ and pixel scale$^{-1}$) resulting from the underlying potential perturbations $\delta\psi_{\rm{GRF}}(\textbf{\textit{x}})$. Consequently, the exclusion probability is nearly insensitive to the slope of the deflection-angle power spectrum $P_{\delta\alpha}(k)$, i.e. the distribution of $\sigma^2_{\delta\alpha}$ over the different length scales. Based on our analysis, $\sigma^2_{\delta\alpha}<6\times10^{-3}$ within the entire considered range $3\leq\beta\leq8$. This insight is valuable for our future analysis of further galaxy-scale lens systems, which might be carried out by perturbing the deflection angle $\alpha(\textbf{\textit{x}})$ instead of the lensing potential $\psi(\textbf{\textit{x}})$. The threshold value itself, however, might vary for different lens systems and depend on the chosen field-of-view, the PSF and the signal-to-noise ratio of the analysed image.
\vspace{0.1cm}


\end{enumerate}

In future work, we intend to investigate the modelling degeneracies in more detail, analyze a larger sample of lens systems, and infer more stringent constraints on the dark-matter and galaxy-formation models by comparing these results to hydrodynamical simulations.

\section*{Acknowledgements}

The authors would like to thank the anonymous reviewer for his/her constructive and valuable comments on this work. We also thank Georgios Vernardos for his suggestions, many of which were very helpful. This study is based on observations made with the NASA/ESA \textit{Hubble Space Telescope}, obtained from the data archive at the Space Telescope Science Institute. Support for this work was provided by a VICI grant (project number 614.001.206) from the Netherlands Organization for Scientific Research (NWO) and by a NASA grant (HST-GO-12898) from the Space Telescope Science Institute. STScI is operated by the Association of Universities for Research in Astronomy, Inc. under NASA contract NAS 5-26555. DB acknowledges support by the Australian Research Council Centre of Excellence for All Sky Astrophysics in 3 Dimensions (ASTRO 3D), through project number
CE170100013. TT acknowledges support by the Packard Foundation through a Packard Research Fellowship. CDF acknowledges support from the NSF under grant AST-1715611. 

\section*{Data availability}
The images and mock data analysed in this work are available from the corresponding author upon reasonable request. The raw HST images are publicly available in the Mikulski Archive for Space Telescopes (MAST).



\bibliographystyle{mnras}
\bibliography{bibliography}



\appendix

\section{Dust analysis}
\label{Appendix:Dust}

Among all HST images available for SDSS J0252+0039 (U-, V-, I- and H-bands), the U-band photometry is most sensitive to the presence of dust. Thus, clumpy dust in the lens galaxy could potentially cause the small-scale variations in the surface brightness of the lensed images that have been so far interpreted as arising from density fluctuations in the lensing-mass distribution. In order to investigate this possible degeneracy, we compare co-aligned images of SDSS J0252+0039 in the U- and I-band. For this comparison, the I-band image is drizzled to the reference frame and pixel size of the U-band image. The U-band image, on the other hand, is smoothed with a Gaussian profile to a larger blur (according to the Gaussian cascade smoothing) of the I-band image: the PSF of the F390W-filter is well described by a Gaussian profile with the FWHM $0.07$ arcsec, whereas the PSF of the F814W filter has the FWHM $0.1$ arcsec.
Subsequently, both images are divided by the respective standard deviation of the photo-electron counts in the empty sky ($0.0012 \, \mathrm{e^-} \, \mathrm{sec^{-1}}$ for the smoothed U-band and $0.0045 \, \mathrm{e^-} \, \mathrm{sec^{-1}}$ for the I-band image drizzled to the pixel scale of the U-band) to obtain the signal-to-noise ratio for each pixel. Finally, to assess the effect of dust, we generate a ratio image between the I- and the U-band and conclude that the variations present across the lensed images can be attributed to differences in the corresponding point-spread functions of the compared filters. An additional visual inspection of the colour image in Fig.~\ref{fig:HST_combined}, based on observations in the UV (F390W), the visual (F814W) and the infrared (F160W) bands, shows no indication for dust extinction either in the lens- or the source galaxy.
 
 

\bsp	
\label{lastpage}
\end{document}